\documentclass[10pt]{iopart}
\usepackage[utf8]{inputenc}
\usepackage{fullpage}

\expandafter\let\csname equation*\endcsname=\relax 
\expandafter\let\csname endequation*\endcsname=\relax 
\usepackage{amsmath}
\usepackage{tensor}
\usepackage{mathtools}
\usepackage{amsfonts}
\usepackage{amssymb}
\usepackage{comment}
\usepackage{tikz}
\usetikzlibrary{calc,positioning}
\usepackage{float}
\usepackage[makeroom]{cancel}
\usepackage{hyperref}
\usetikzlibrary{decorations.pathmorphing}
\usetikzlibrary{decorations.pathreplacing}
\allowdisplaybreaks

\newcommand{\drm}{\mathrm{d}} 
\newcommand{\h}{\mathcal{H}} 
\newcommand{\J}{\mathcal{J}} 
\newcommand{\s}{\Sigma} 
\newcommand{\B}{\Sigma'} 
\newcommand{\M}{\mathcal{M}} 
\newcommand{\C}{\mathcal{C}} 

\newcommand{\dd}{\mathrm{d}}
\newcommand{\beq}{\begin{equation}}
\newcommand{\eeq}{\end{equation}}

\def\al{\alpha}
\def\be{\beta}
\def\la{\lambda}

\def\a{a}
\def\b{b}
\def\c{c}
\def\d{d}
\def\ta{\tilde{\lambda}}

\tikzset{snake it/.style={decorate, decoration=snake}}
\definecolor{lightgreen}{rgb}{0.56, 0.93, 0.56}

\begin{document}

\title{On the horizon entropy of a causal set}

\author{Ludovico Machet$^1$$^,$$^2$ and Jinzhao Wang$^3$}
\address{$^1$ Institute for Theoretical Physics, KU Leuven, Celestijnenlaan 200D, B-3001 Leuven, Belgium}
\address{$^2$ Theoretical and Mathematical Physics, ULB, Boulevard du Triomphe, B-1050 Bruxelles, Belgium}
\address{$^3$ Institute for Theoretical Physics, ETH, 8093 Z\"urich, Switzerland}

\ead{ludovico.machet@kuleuven.be, jinzwang@phys.ethz.ch}

\begin{abstract}
We discuss how to define a kinematical horizon entropy on a causal set. We extend a recent definition of horizon molecules to a setting with a null hypersurface crossing the horizon. We argue that, as opposed to the spacelike case, this extension fails to yield an entropy local to the hypersurface-horizon intersection in the continuum limit when the causal set approximates a curved spacetime. We then investigate the entropy defined via the Spacetime Mutual Information between two regions of a causal diamond truncated by a causal horizon, and find it does limit to the area of the intersection.
\end{abstract}

\section{Introduction}
\label{intro}

In the last decades, the claim that thermodynamics-like laws apply to Black Holes has accumulated an ever-growing amount of evidence. After the pioneering work of Hawking \cite{Hawking} and Bekenstein \cite{Bekenstein} that established the Black Hole thermodynamic laws, those properties have been generalised to cosmological horizons and they are now believed to be valid for all causal horizons \cite{JP}. A quantum field theoretical approach to the problem, together with the finiteness of Black Hole entropy, suggest spacetime must exhibit a discrete structure at Planck scales. Thus, one can think of the thermodynamics of causal horizons as emerging from the dynamics of some microscopical degree of freedom, each of which roughly of the size of the Planck scale and carrying one bit of entropy. \\

Causal Set Theory (CST) offers a naturally discrete setup through which one can gain insights on these subjects. Spacetime is postulated to be a locally finite and partially ordered set, with the partial order encoding causality relations between the set elements, i.e. spacetime events. One call such a set a causal set, or causet. The continuum-like geometry usually invoked to describe gravity would then be a coarse grained average over a statistical ensemble of causets \cite{Surya}. The intuition that classical thermodynamics hints towards the molecular nature of matter motivates studying horizon thermodynamics to look for signatures of the fine grained structure of spacetime. Multiple attempts have been made to define a kinematical entropy on a causal set. One way was to equal the number of causal links crossing the horizon to the horizon entropy, which gave rise to the \textit{horizon molecules} program. Admittedly, horizon molecule can only be a coarse-grained notion of entropy, but it might offer a hint to a deeper understanding of gravitational entropy in the context of causal set theory or generally other discretized theories. See more motivations for studying horizon molecules in \cite{DS,Marr,Dowker,Surya}. There is another more fine-grained program that studied entropy for quantum fields defined on a causal set by studying correlation functions between spacetime regions. We shall not comment on this line of research but it is an interesting open problem to draw links between the two approaches \cite{Sorkin_2018,chen2020towards,surya2020entanglement}. \\

In section \ref{horizon molecules}, we will quickly review the horizon molecules proposal, recalling the original definition due to Dou and Sorkin in 2D. Their proposal correctly produces the area law entropy given by the intersection of the horizon with any null hypersurface. However, the drawback is that it cannot be applied to higher dimensions due to divergence issues. Their proposal was recently refined by Barton et al and applied to all spacetime dimensions concerning any spacelike hypersurface $\s$ crossing the horizon $\h$. Their main result is the following claim: \emph{in the continuum limit, the expected number of horizon molecules is equal
to the area of the horizon intersection with a spacelike hypersurface, $\J:=\s\cap\h$, in discreteness units, up to a dimension dependent constant of order one.} 
 \begin{equation}
 	\lim_{\rho\rightarrow\infty}\rho^{\frac{d-2}{d}}\langle \mathbf{H}(\s)\rangle=a^{(d)}\int_\J\drm V_\J, 
 	\label{claim_Dowker}
 \end{equation}
where $\rho$ denotes the causet density and $\langle \mathbf{H}(\s)\rangle$ is the expectated number of horizon molecules with respect to the Poisson sprinkling on the region bounded by $\s$ (see details in section \ref{horizon molecules}). \\

 Note that the restriction here that $\s$ being spacelike is important to the arguments of Barton et al. One might naively expect that since (\ref{claim_Dowker}) holds for \emph{any} spacelike hypersurfaces, it should also hold for any null hypersurface $\s$, which can be casted as a limit of a sequence of spacelike hypersurfaces $\s=\lim_{t\rightarrow\infty} \s_t$. We can construct an one-parameter family of spacelike hypersurfaces $\s_t$ that continuously deform to the null hypersurface $\s$ as we pass to the limit.  This is promising as the region we sprinkle into and so the probability measure is also continuous with respect to the deformations. However, we need to be extra careful here because we are also taking the infinite density limit, and it is not a priori true that the limits can be exchanged. We would like to have
 \begin{equation}
      \lim_{\rho\rightarrow\infty}\rho^{\frac{d-2}{d}}\langle \mathbf{H}(\s)\rangle=\lim_{\rho\rightarrow\infty}\lim_{t\rightarrow\infty}\rho^{\frac{d-2}{d}}\langle \mathbf{H}(\s_t)\rangle \stackrel{?}{=} \lim_{t\rightarrow\infty}\lim_{\rho\rightarrow\infty}\rho^{\frac{d-2}{d}}\langle \mathbf{H}(\s_t)\rangle = \lim_{t\rightarrow\infty} a^{(d)}\int_{\J_t}\drm V_{\J_t} = a^{(d)}\int_{\J}\drm V_{\J}.
 \end{equation}
where $\J_t:=\s_t\cap\h$ and we've applied (\ref{claim_Dowker}) and the continuity of the codimension-two volume in the last two equalities. \\

Since in general the limits do not commute, we shall expect a different limiting behaviour as opposed to (\ref{claim_Dowker}). We will then apply the Barton et al proposal to two specific cases of null hypersurfaces crossing the horizon, one being folded null planes and the other being downward light-cones. In the continuum limit, we recover the same area law in the Minkowski spacetime in section \ref{flat}.  However in section \ref{curved}, we argue that whenever the hypersurface contains a null segment crossing the horizon, there are generically non-local contributions from the null segment to the horizon molecules expectation values in the continuum limit, which is in contrast with a definition of entropy local to the horizon. The area law is therefore distorted.\\

\emph{In the continuum limit, the expected number of horizon molecules is generically not equal
to the area of the horizon intersection $\J$ with a hypersurface $\Sigma$ containing a null segment in the neighbourhood of $\J$ in discreteness units.} 
 \begin{equation}
 	\lim_{\rho\rightarrow\infty}\rho^{\frac{d-2}{d}}\langle \mathbf{H}(\s)\rangle \neq a^{(d)}\int_\J\drm V_\J, 
 	\label{claim_MW}
 \end{equation}
 
We explicitly calculate the deviations in the downward light-cone case example. It uses a causal diamond volume expansion in the limit that the causal diamond shrinks towards a null geodesic. The volume formula might be of independent interest in order problems so we include the details in the Appendix.\\

We don't immediately see a sensible way to adjust the Barton et al proposal such that it recovers the area law for both spacelike or null hypersurfaces.  On the other hand, an alternative definition for horizon entropy is given by the \textit{spacetime mutual information} (SMI) proposal, which looks at the failure of additivity of the causet action between two regions in order to infer the entropy of a causal horizon. In section \ref{SMI}, we shall consider the SMI of a causal diamond in Minkowski spacetime sliced through by a Rindler horizon, and show that the SMI between the top and bottom region of the diamond yields the expected area law for the horizon entropy. From this result, we can also infer a non-trivial check of the Benincasa-Dowker (BD) action conjecture on the causet action continuum limit \cite{buck2015boundary}. 

\section{Horizon molecules}
\label{horizon molecules}
 
Dou and Sorkin \cite{DS} proposed the causal relations between causal set's elements are the fundamental structures giving rise to the horizon entropy. In particular, they equated the horizon entropy, defined on a hypersurface $\s$ crossing the horizon, to the number of causal links crossing the horizon in the vicinity of $\s$. Let's consider a spacetime $M$ faithfully approximated by a causal set $\C$.  Let $\h$ be a black hole horizon and let $\s$ be a non-timelike hypersurface crossing the horizon and on which we want to measure the horizon entropy. One can therefore define a \textit{horizon molecule} as \\
 
 \textbf{Definition 0 (Dou-Sorkin) - } The pair $\{x,y\}$ of elements of $\C$ is a horizon molecule if 
 
 \begin{enumerate}
 	\item $y\in I^-(\s)\cap I^-(\h)$,
 	\item $x\in I^+(\s)\cap I^+(\h)$,
 	\item $|I[y,x]|=0$, i.e. $y\prec x$ is a link,
 	\item $y$ is maximal in $I^-(\s)\cap I^-(\h)$ and $x$ is minimal in $I^+(\h)$.
 \end{enumerate}

 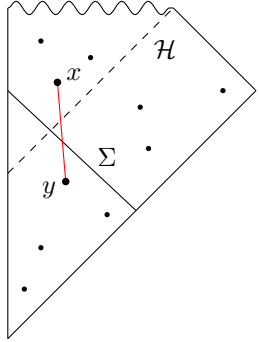
\begin{figure}[h]
 	\centering 
 	\begin{tikzpicture}[scale=1.1]
 		\draw[-] (0,0) -- (0,4);
 		\draw[-,snake it] (0,4) -- (2,4);
 		\draw[-] (0,0) -- (3,3);
 		\draw[-] (3,3) -- (2,4);
 		\draw[-, dashed] (0,2) -- (2,4);
 		\draw[-] (0,3) -- (1.55,1.55);
 		
 		\draw[-,red] (.7,1.9) -- (.6,3.1);
 		\node at (.7,1.9)[circle,fill,inner sep=1pt]{};
 		\node at (.6,3.1)[circle,fill,inner sep=1pt]{};
 		
 		\node at (.8,3.2) {$x$};
 		\node at (.5,1.8) {$y$};
 		\node at (1.2,2.2) {$\Sigma$};
 		\node at (1.9,3.5) {$\mathcal{H}$};
 		
 		\node at (.4,1.1)[circle,fill,inner sep=.7pt]{};
 		\node at (1.2,1.5)[circle,fill,inner sep=.7pt]{};
 		\node at (.2,0.6)[circle,fill,inner sep=.7pt]{};
 		\node at (.4,3.6)[circle,fill,inner sep=.7pt]{};
 		\node at (1,3.4)[circle,fill,inner sep=.7pt]{};
 		\node at (1.7,2.3)[circle,fill,inner sep=.7pt]{};
 		\node at (1.6,2.8)[circle,fill,inner sep=.7pt]{};
 		\node at (2.6,3)[circle,fill,inner sep=.7pt]{};
 		
 	\end{tikzpicture}
 	\caption{A Dou-Sorkin horizon molecule in a black hole spacetime.}
 	\label{DS_Hm}
 \end{figure}
 
An intuition of this definition is given in figure \ref{DS_Hm}. The expected number of horizon molecules defined as such was shown to be proportional to the horizon area in the case of a $2$-dimensional black hole spacetime. This promising result was soon undercut by the realisation that this definition yields a pathological divergence in $d>2$. The intersection of the horizon with the future light-cone of $y$ is an unbounded surface and elements $x$ are allowed to be sampled arbitrarily far away along this intersection. This IR divergence issue was pointed out in \cite{Marr}, where new definitions were introduced to cure this divergence. However, that make the mathematical handling of the subject lengthy and cumbersome.\\  
 
Recent work from Barton et al \cite{Dowker} revisited the Dou-Sorkin proposal and extended the definition to more general causal horizons. The idea is to modify \textbf{Definition 0} and require the hypersurface $\s$ to be strictly spacelike. The definition goes as follows. Let $(M,g)$ be a globally hyperbolic spacetime with a Cauchy surface $\s$. Let $\h$ be a causal horizon, defined as the boundary of the past of a future inextendible timelike curve $\gamma$, i.e. $\h:=\partial I^-(\gamma)$, and consider a random causal set $(\C,\prec)$ generated on $M$ through a Poisson sprinkling.\\
 
 \textbf{Definition 1  (Barton et al) - } A horizon molecule is a pair of elements of $\C$, $\{p_-, p_+\}$, such that 
 
 \begin{enumerate}
 	\item $p_-\prec p_+$,
 	\item $p_-\in I^-(\s)\cap I^-(\h)$,
 	\item $p_+\in I^-(\s)\cap I^+(\h)$,
 	\item $p_+$ is the only element in $I^-(\s)$ and in the future of $p_-$. 
 \end{enumerate}
 
\begin{figure}[h]
 	\centering 
 	\begin{tikzpicture}[scale=.9]
 		\draw [-] plot [smooth, tension=1] coordinates { (-2.5,-.2) (0,0.1) (3,-.2) (4.5,.1)};
 		\draw[-] (-.5,-2) -- (3.4,1.9);
 		
 		\draw[-, dashed] (1.1,-1.2) -- (2.12,-0.12);
 		\draw[-, dashed] (1.1,-1.2) -- (-0.4,0.1);
 		
 		\draw[-,red] (1.1,-1.2) -- (.8,-.25);
 		
 		\node at (1.1,-1.2)[circle,fill,inner sep=1pt]{};
 		\node at (.8,-.25)[circle,fill,inner sep=1pt]{};
 		
 		\node at (1.4,-1.4) {$p_-$};
 		\node at (0.5,-0.2) {$p_+$};
 		\node at (-1.8,.3) {$\Sigma$};
 		\node at (2.6,1.6) {$\mathcal{H}$};
 		\node at (1.4,.4) {$\mathcal{J}$};
 		
 		\node at (.4,1.1)[circle,fill,inner sep=.7pt]{};
 		\node at (1.2,1.5)[circle,fill,inner sep=.7pt]{};
 		\node at (.2,0.6)[circle,fill,inner sep=.7pt]{};
 		\node at (2.8,.6)[circle,fill,inner sep=.7pt]{};
 		\node at (-1,-.5)[circle,fill,inner sep=.7pt]{};
 		\node at (-1.7,-1.3)[circle,fill,inner sep=.7pt]{};
 		\node at (1.9,-.8)[circle,fill,inner sep=.7pt]{};
 		\node at (2.5,-1.3)[circle,fill,inner sep=.7pt]{};
 		\node at (3.5,.2)[circle,fill,inner sep=.7pt]{};
 		\node at (-1,1)[circle,fill,inner sep=.7pt]{};
 		
 	\end{tikzpicture}
 	\caption{Barton et al definition of horizon molecules with respect to a spacelike hypersurface $\s$.}
 	\label{D-Hm}
 \end{figure}
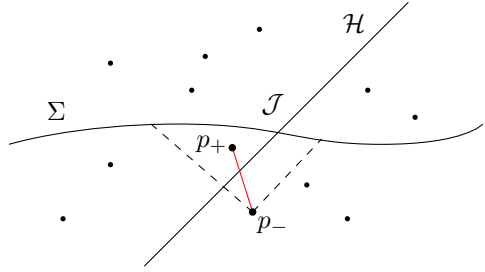

We give a graphical intuition on figure \ref{D-Hm}. This definition can be extended to a $n$-molecule $\{p_-,p_{+,1},...,p_{+,n}\}$ by requiring $p_-\prec p_{+,k}$ and $\{p_{+,1},...,p_{+,n}\}$ to be the only elements in the future of $p_-$ and in $I^-(\s)$. We are going to use \textbf{Definition 1} for the rest of this paper. It was shown that in the continuum limit, the expected number of such defined horizon molecules limits to the area of the intersection of $\s$ and $\h$ up to a constant dependent on the spacetime dimension. Mathematically, we restate (\ref{claim_Dowker}) here:

 \begin{equation}
 	\lim_{\rho\rightarrow\infty}\rho^{\frac{d-2}{d}}\langle \mathbf{H}\rangle=a^{(d)}\int_\J\drm V_\J, 
 \end{equation}
 
 where $\mathbf{H}$ is the random variable counting the number of horizon molecules in the sprinkling realisations, $\J=\s\cap\h$ and $\drm V_\J$ the surface measure on $\J$. \\
 
\subsection{Horizon molecule counting with null hypersurfaces}

\begin{figure}[h]
	\centering 
	\begin{tikzpicture}[scale=1]
		\draw[-] (-1.5,-1.5)--(1.7,1.7);
		\draw[-] (2,-1)-- (-0.25,1.25);
		\draw[-] (-0.25,1.25)-- (-2,-.5);
		
		\node at (2.1,-0.6) {$\Sigma$};
		\node at (1.85,1.35) {$\mathcal{H}$};
		\node at (.85,.5) {$\mathcal{J}$};
		
		\node at (-.1,1.6)[circle,fill,inner sep=.7pt]{};
		\node at (.4,1.1)[circle,fill,inner sep=.7pt]{};
		\node at (1.2,1.5)[circle,fill,inner sep=.7pt]{};
		\node at (-1,-.5)[circle,fill,inner sep=.7pt]{};
		\node at (-1.7,-1.3)[circle,fill,inner sep=.7pt]{};
		\node at (1,-.8)[circle,fill,inner sep=.7pt]{};
		\node at (2,0)[circle,fill,inner sep=.7pt]{};
		\node at (1.4,.6)[circle,fill,inner sep=.7pt]{};
		\node at (-1,1.2)[circle,fill,inner sep=.7pt]{};
		\node at (-1.8,.8)[circle,fill,inner sep=.7pt]{};
		\node at (-.5,-1.3)[circle,fill,inner sep=.7pt]{};
		
		\draw[red] (.2,-0.3) -- (-0.1,.6);
		
		\node at (.2,-0.3)[circle,fill,inner sep=1pt]{};
		\node at (-0.1,.6)[circle,fill,inner sep=1pt]{};
		
		\draw[dashed] (.2,-.3) -- (.75,.23);
		\draw[dashed] (.2,-.3) -- (-.77,.72);
		
		\node at (0,-0.55) {$p_-$};
		\node at (-0.1,.8) {$p_+$};
		
	\end{tikzpicture}
	\caption{We define the horizon molecule to be constrained into the past of $\s$.}
	\label{M-Hm}
\end{figure}
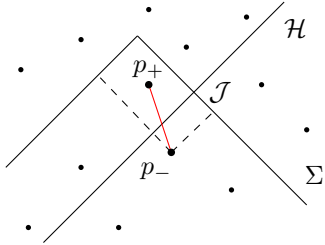

We now discuss the possibility to apply the Barton et al definition to a null hypersurface $\s$. To avoid the IR divergence issues of Dou-Sorkin, we cannot use a straight null plane that renders the sprinkling domain unbounded. Instead, we shall put some regulators to bound the domain, or better to have it included as part of the null hypersurface.  It's natural to consider the case of $\s$ being a folded null plane and a downward light-cone respectively. The idea comes from the formulation of Einstein field equations solutions in term of a characteristic Cauchy problem \cite{Cauchy}. Let us consider a spacetime $(\mathcal{M},g)$ and a null submanifold $\h$ defining a causal horizon. Let us then consider a null hypersurface $\s$ transverse to the horizon with a kink in the future of $\h$. Call $\J$ the intersection $\J:=\h\cap\s$. Let us assume a causal set $\C$ is sampled on $(\mathcal{M},g)$ through a Poisson sprinkling at density $\rho$. Then, the definition of horizon molecules given by Barton et al still holds and we are left with the setup of figure \ref{M-Hm}.

%\textbf{Definition 1. } A horizon molecule is a pair of elements of $\C$, $\{p_-,p_+\}$, such that
%\begin{enumerate}
%	\item $p_-\prec p_+$,
%	\item $p_-\in I^-(\s)\cap I^-(\h)$,
%	\item $p_+\in I^-(\s)\cap I^-(\B)\cap  I^+(\h)$,
%	\item $p_+$ is the only element in both $I^-(\s)$ and in the future of $p_-$.
%\end{enumerate}

%Thus, a horizon molecule is a link, Figure \ref{M-Hm}. Analogously, one can define\\

%\textbf{Definition 2 } A horizon $n$-molecule is a subcauset of $\mathcal{C}$, $\{p_-,p_{+,1},...,p_{+,n}\}$, such that
%\begin{enumerate}
%	\item $p_-\prec p_{+,k}$, for all $k=1,...,n$,
%	\item $p_-\in I^-(\s)\cap I^-(\h)$,
%	\item $p_{+,k}\in I^-(\s)\cap I^-(\B)\cap  I^+(\h)$,  for all $k=1,...,n$,
%	\item $\{p_{+,1},...,p_{+,n}\}$ are the only elements in both $I^-(\s)$ and in the future of $p_-$.
%\end{enumerate}

Defining now the expectation value of the number of horizon $n$-molecules as in \cite{Dowker}, and calling $p$ the element $p_-$, one gets to the following identity

\begin{equation}
	\rho^{\frac{2-d}{d}}\big\langle\mathbf{H}_n\big\rangle=\rho^{\frac{2-d}{d}+1}\int_{I^-(\J)}\,\drm V_p \frac{(\rho V_+(p))^n}{n!}e^{-\rho V(p)},
	\label{entropy_def}
\end{equation}

where the volumes entering the integral are defined as
\begin{equation}
    \begin{aligned}
	V_+(p):=&\mathrm{vol}\big( I^-(\s)\cap  I^+(\h) \cap I^+(p)\big),\\
	V(p):=&\mathrm{vol}\big( I^-(\s) \cap I^+(p)\big).
\end{aligned}
\end{equation}

In order to make the computations more straightforward, it is useful to revert the time direction of our system, so that one lets $p$ run in the future of the intersection hypersurface $\J$. Equation \eqref{entropy_def} takes the form 

\begin{equation}
	\rho^{\frac{2-d}{d}}\big\langle\mathbf{H}_n\big\rangle=\rho^{\frac{2-d}{d}+1}\int_{I^+(\J)}\,\drm V_p \frac{(\rho V_+(p))^n}{n!}e^{-\rho V(p)}.
	\label{entropy_def_b}
\end{equation}

\section{Minkowski spacetime}
\label{flat}

We will now probe our definition on a Minkowski $d$-dimensional spacetime. Let's consider a $d$-dimensional Minkowski  spacetime $(\mathbb{M}^d,\eta)$ with a submanifold $\mathcal{H}$ defining a Rindler causal horizon. We're going to compute the expected number of horizon molecules with two distinct null hypersurface configurations, and find that they both agree with the area-law entropy.

\subsection{Folded null planes}
Let $\s$ be a null hypersurface transverse to the horizon and intersecting it in a co-dimension $2$ surface $\mathcal{J}=\s\cap\h$. We set up the coordinates $(v,u,y^\alpha)$ and let $\s$ be the $v=0$ hypersurface and $\h$ the $u=0$ one. We define the hypersurface $\B$ given by $u=\lambda$, with $\lambda>0$, which then lies in the future of $\h$ and it is parallel to the horizon. The union $\B\cup\s$ form a $\wedge$-shaped folded null plane, and we can ignore the leftover parts on the top. \\

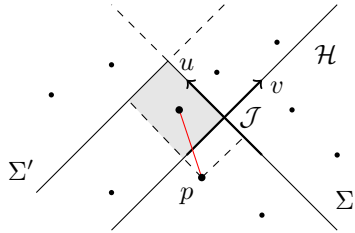
\begin{figure}[h]
	\centering 
	\begin{tikzpicture}[scale=1]
	
	    \fill[gray!20] (-.25,1.25) -- (-.77,.72) -- (-0.05,-0.05) -- (0.5,.5)-- cycle;
		\draw[-] (-1,-1)--(2,2);
		\draw[-] (2,-1)-- (-.25,1.25);
		\draw[-, dashed] (-.25,1.25)-- (-1,2);
		\draw[-] (-.25,1.25)-- (-2,-.5);
		\draw[-, dashed] (0.5,2)-- (-.25,1.25);
		
		\node at (-2.2,-0.2) {$\Sigma'$};
		\node at (2.1,-0.6) {$\Sigma$};
		\node at (1.85,1.35) {$\mathcal{H}$};
		\node at (.85,.5) {$\mathcal{J}$};
		
		\draw[->,line width=0.3mm] (0,0)--(1,1);
		\node at (1.2,.9) {$v$};
		\draw[->,line width=0.3mm] (1,0)--(0,1);
		\node at (0,1.2) {$u$};
		
		\node at (.4,1.1)[circle,fill,inner sep=.7pt]{};
		\node at (1.2,1.5)[circle,fill,inner sep=.7pt]{};
		\node at (-1,-.5)[circle,fill,inner sep=.7pt]{};
		\node at (1,-.8)[circle,fill,inner sep=.7pt]{};
		\node at (2,0)[circle,fill,inner sep=.7pt]{};
		\node at (1.4,.6)[circle,fill,inner sep=.7pt]{};
		\node at (-1,1.2)[circle,fill,inner sep=.7pt]{};
		\node at (-1.8,.8)[circle,fill,inner sep=.7pt]{};
		
		\draw[red] (.2,-0.3) -- (-0.1,.6);
		
		\node at (.2,-0.3)[circle,fill,inner sep=1pt]{};
		\node at (-0.1,.6)[circle,fill,inner sep=1pt]{};
		
		\draw[dashed] (.2,-.3) -- (.75,.23);
		\draw[dashed] (.2,-.3) -- (-.77,.72);
		
		\node at (0,-0.55) {$p$};
		
	\end{tikzpicture}
	\caption{A folded null plane crossing the horizon. The volume $V_+,f$ is shaded}
	\label{Hm-Minkowski_1}
\end{figure}

In this setup, the integrand of equation \eqref{entropy_def_b} is independent of $y^\alpha$. Thus one can write

\begin{equation}
	\rho^{\frac{2-d}{d}}\big\langle\mathbf{H}_n\big\rangle=\int_{\J}\,\drm^{d-2}y\,I_n^{(d,\mathrm{flat})}(l,\lambda), 
	\label{entropy_def_c}
\end{equation}

with the function $I_n^{(d,\mathrm{flat})}(l,\lambda)$  given by

\begin{equation}
	I_n^{(d,\mathrm{flat})}(l,\lambda)=\frac{l^{-(dn+2)}}{n!}\int_0^\infty\drm v\int_0^\infty\drm u\,\left(V_{+,f}(u,v,\lambda)\right)^ne^{-l^{-d} V_f(u,v,\lambda)},
	\label{I_flat}
\end{equation}

Where we denoted with a $f$ subscript the quantities evaluated on the Minkowski spacetime. As already said, for convenience we will work with a reversed time direction, so that $p\in I^+(\J)$. Then, the surface $\B$ will be given by $u=-\lambda$. One can compute the volumes $V_f$ and $V_{+,f}$ in a coordinate system $(x^0,x^1,y^\alpha)$ such that

\begin{align}
	v=\frac{x^0+x^1}{\sqrt{2}},\quad\quad u=\frac{x^0-x^1}{\sqrt{2}}, 
\end{align} 

and so that the point $p$ as $x^1=0$. Thus, $V_{-,f}:=V_f-V_{+,f}$ is given by the difference between the volume of a cone of height $x_p^0$ and the volumes of two identical regions between the surfaces $\Sigma$ (and $\h$) and a spacelike surface $x^0=0$. One has 

\begin{equation}
	V_{-,f}=\frac{\Omega_{d-2}}{d(d-1)}(x^0_p)^d - 2 \int_0^{\frac{1}{2}x^0_p}\drm x^{'0}\int_{x^{'0}-x^0}^{x^0}\drm x^{'1}\int_0^{\sqrt{(x^0-x^{'0})^2-(x^{'1})^2}}\drm R\int_{\mathbb{S}_{d-3}}\drm\Omega_{d-3}\,R^{d-3}=\alpha_d(x^0_p)^d,
\end{equation}

where $\alpha_d$ is a constant dependent on the dimension. This suggests volumes of the form of $V_{-,f}$ and $V_f$ are dependent only on the proper time between the origin and the point $p$. Thus, one gets

\begin{align}
	V_f(u,v,\lambda)&=\alpha_d\left(2v(u+\lambda)\right)^{\frac{d}{2}},\label{flat_va}\\
	V_{+,f}(u,v,\lambda)&=\alpha_d(2v)^{\frac{d}{2}}\left((u+\lambda)^{\frac{d}{2}}-u^{\frac{d}{2}}\right).
	\label{flat_vb}
\end{align}

Substituting equations \eqref{flat_va} and \eqref{flat_vb} into \eqref{I_flat} yields the result

\begin{align}
	I_n^{(d,\mathrm{flat})}(l,\lambda)=&\frac{l^{-(dn+2)}}{n!}\int_0^\infty\drm u\int_0^\infty\drm v\,\alpha_d^n(2v)^{\frac{dn}{2}}\left((u+\lambda)^{\frac{d}{2}}-u^{\frac{d}{2}}\right)^ne^{-l^{-d} \alpha_d\left(2v(u+\lambda)\right)^{\frac{d}{2}}},\nonumber\\
	=& \frac{l^{-(dn+2)}}{n!}\int_0^\infty\drm u\,\frac{\alpha_d^{-\frac{2}{d}}}{d}l^{dn+2}\left((u+\lambda)^{\frac{d}{2}}-u^{\frac{d}{2}}\right)^n(u+\lambda)^{-1-\frac{dn}{2}}\Gamma\left[\frac{2}{d}+n\right],\nonumber\\
	=& \frac{\alpha_d^{-\frac{2}{d}}}{n!d}\Gamma\left[\frac{2}{d}+n\right]\int_0^\infty\drm u\,\left((u+\lambda)^{\frac{d}{2}}-u^{\frac{d}{2}}\right)^n(u+\lambda)^{-1-\frac{dn}{2}}.
\end{align}

One can now evaluate the integral for different values of $n$, for example

\begin{align}
	I_1^{(d,\mathrm{flat})}=a_1^{(d)}=&\frac{\alpha_d^{-\frac{2}{d}}}{d}\Gamma\left[\frac{d+2}{d}\right]H_{\frac{d}{2}},\nonumber\\
	I_2^{(d,\mathrm{flat})}=a_2^{(d)}=&\frac{\alpha_d^{-\frac{2}{d}}}{2d}\Gamma\left[2+\frac{2}{d}\right]\left(2H_{\frac{d}{2}}-H_d\right),
\end{align}

with $H_n$ the $n^\mathrm{th}$ harmonic number. We notice the result is independent of the $\B$ hypersurface position with respect to the horizon, i.e. of the parameter $\lambda$. It follows that 

\begin{equation}
	\lim_{\rho\rightarrow\infty}\rho^{\frac{2-d}{d}}\big\langle\mathbf{H}_n\big\rangle=a_n^{(d)}\int_{\J}\,\drm V_\J
	\label{claim_M}
\end{equation}

holds in $\mathbb{M}_d$. This result is surprising as one would expect the locality argument given in Barton et al to fail in this setup. This is because in $\eqref{entropy_def}$ the term $\mathrm{exp}(-\rho V)$ suppresses the contributions of points for which $\rho V\gg 1$, thus the points far from the intersection $\J$ in the case $\s$ is spacelike. With $\s$ null, contributions from points sampled arbitrarily close to $\s$ and arbitrarily far away from $\J$ will not be exponentially suppressed in the continuum limit. However, because of dimensional analysis, one can argue that the continuum limit of \eqref{entropy_def_c} must be dependent only of geometrical quantities at $\J$. This is because of the symmetries of flat spacetime. One can always boost the system in the $u$ direction to pull the surface $\B$ arbitrarily close to $\h$, thus the independence on $\lambda$, or points far in the past of $\J$ closer to the intersection.

\subsection{Downward Light-cone}
\label{LC_f}

Now we consider $\s$ given by a downward light-cone, that is the boundary of the past of a point $q\in I^+(\h)$. Then the horizon molecules will be sprinkled in the causal interval defined by the points $p$ and $q$, i.e. $I{[p,q]}$, see figure \ref{Hm-Minkowski_2}. Equation \eqref{entropy_def} remains the same, with the volumes given by

\begin{equation}
 \begin{aligned}
	V_+(p):=&\mathrm{vol}\big( I{[p,q]}\cap   I^+(\h) \big),\\
	V(p):=&\mathrm{vol}\big( I{[p,q]}).
\end{aligned}   
\end{equation}

%\textbf{Definition 3} A horizon $n$-molecule is a subcauset of $\mathcal{C}$, $\{p,p_{+,1},...,p_{+,n}\}$, such that
%\begin{enumerate}
%	\item $p\in I^-(q)\cap I^-(\h)$,
%	\item $p_{+,k}\in I{[p,q]}\cap  I^+(\h)$,  for all $k=1,...,n$,
%	\item $\{p_{+,1},...,p_{+,n}\}$ are the only elements in $I{[p,q]}$.
%\end{enumerate}

\iffalse
\begin{figure}[h]
	\centering 
	\begin{tikzpicture}[scale=1.2]
		\draw[-] (-1,-1)--(1.5,1.5);
		\draw[-] (2,-1)-- (-0.25,1.25);
		\draw[-] (-0.25,1.25)-- (-2,-.5);
		
		\node at (-2.5,-0.2) {$\partial I^-(q)$};
		\node at (1,1.4) {$\mathcal{H}$};
		\node at (.85,.5) {$\mathcal{J}$};
		
		\node at (.4,1.1)[circle,fill,inner sep=.7pt]{};
		\node at (-1,-.5)[circle,fill,inner sep=.7pt]{};
		\node at (1,-.8)[circle,fill,inner sep=.7pt]{};
		\node at (2,0)[circle,fill,inner sep=.7pt]{};
		\node at (1.4,.6)[circle,fill,inner sep=.7pt]{};
		\node at (-1,1.2)[circle,fill,inner sep=.7pt]{};
		\node at (-1.8,.8)[circle,fill,inner sep=.7pt]{};
		
		\draw[red] (.2,-0.3) -- (-0.1,.6);
		
		\node at (.2,-0.3)[circle,fill,inner sep=1pt]{};
		\node at (-0.1,.6)[circle,fill,inner sep=1pt]{};
		\node at (-.25,1.25)[circle,fill,inner sep=1pt]{};
		
		\draw[dashed] (.2,-.3) -- (.75,.23);
		\draw[dashed] (.2,-.3) -- (-.77,.72);
		
		\node at (0,-0.55) {$p$};
		\node at (-0.3,1.55) {$q$};
		
	\end{tikzpicture}
	\caption{Two-dimensional sketch of a horizon molecules defined with respect to a downward light-cone.}
	\label{Hm-Minkowski_2}
\end{figure}
\fi

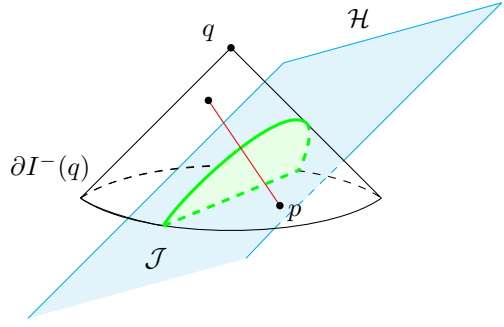
\begin{figure}[h]
	\centering 
	\begin{tikzpicture}
		
		\fill[cyan!10] (.2,-.8) -- (3.6,2.6) -- (0.7,1.8) -- (-2.7,-1.6)-- cycle;
		\draw[cyan] (3.6,2.6) -- (0.7,1.8) -- (-2.7,-1.6);
		\draw[cyan] (3.6,2.6) -- (1.5,.5);
		\draw[cyan, dashed] (0.58,-.4) -- (1.5,.5);
	    \draw[cyan] (0.58,-.4) -- (0.2,-.8);
		
		\draw[-] (2,0) -- (0,2) -- (-2,0) ;
		
		\node at (0,2)[circle,fill,inner sep=1pt]{};
		\node at (-0.3,2.2) {$q$};
		
		\node at (1.7,2.4) {$\mathcal{H}$};
		\node at (-1,-.8) {$\mathcal{J}$};
		\node at (-2.4,0.4) {$\partial I^-(q)$};
		
		\draw (-2,0) to[out=-35,in=-145, distance=1cm ] (2,0);
		\draw[dashed] (-2,0) to[out=35,in=-215, distance=1cm ] (2,0);
		
		\draw (-2,0) to[out=-35,in=-190, distance=.3cm ] (-.9,-.37);
		\draw[dashed] (-2,0) to[out=35,in=-190, distance=.8cm ] (.9,.37);
		
		\fill[green!10]  (1,1) to[out=140,in=60, distance=.45cm] (-.9,-.37) -- (.9,.37) to[in=320,out=60, distance=.15cm] (1,1);
		
		\draw[line width=0.4mm, green] (1,1) to[out=140,in=60, distance=.45cm] (-.9,-.37);
		\draw[line width=0.4mm, dashed, green] (1,1) to[out=320,in=60, distance=.15cm] (.9,.37);
		\draw[line width=0.4mm, dashed, green] (-.9,-.37) -- (.9,.37);

		\draw[red] (0.65,-0.1) -- (-0.3,1.3);
		\node at (0.65,-0.1)[circle,fill,inner sep=1pt]{};
		\node at (-0.3,1.3)[circle,fill,inner sep=1pt]{};
		\node at (0.85,-0.2) {$p$};
		
	\end{tikzpicture}
	\caption{Sketch of a horizon molecules defined with respect to a downward light-cone.}
	\label{Hm-Minkowski_2}
\end{figure}

We claim equation \eqref{claim_M} holds in this configuration too. To supply evidence for this result, consider a causal diamond $I{[p,q]}$ as defined above of proper length $\tau$. In analogy to the previous calculations, suppose the time direction to be swapped, so that $p\in I^+(\h)$ and $q\in I^-(\h)$. By Lorentz invariance, one can compute its volume in a frame in which de diamond is spherically symmetric. Again we set up the coordinates system $(x^0,x^1,y^\alpha)$ such that the origin sits in the middle of the diamond and we choose 

\begin{align}
	p^\mu&=\left(\frac{\tau}{2},0,...,0\right), & q^\mu&=\left(-\frac{\tau}{2},0,...,0\right).
\end{align}

Suppose then the horizon $\h$ is a Rindler horizon given by the condition $x^0=x^1+\alpha$, with $\alpha\in(-\frac{\tau}{2},\frac{\tau}{2})$. Therefore, one can write the volume of the region $ I_{[p,q]}\cap   I^-(\h) $ as 

\begin{equation}
  \begin{aligned}
	V_{+,f}^{(d,\alpha)}&= \int_{-\frac{\tau}{2}}^{-\frac{1}{2}\left(\frac{\tau}{2}-\alpha\right)}\drm x^0\int_{-x^0-\frac{\tau}{2}}^{x^0+\frac{\tau}{2}}\drm x^1\int_0^{\sqrt{\left(x^0+\frac{\tau}{2}\right)^2-(x^1)^2}}\drm R\int_{\mathbb{S}_{d-3}}\drm\Omega_{d-3}\,R^{d-3}\\
	&+\int_{-\frac{1}{2}\left(\frac{\tau}{2}-\alpha\right)}^{0}\drm x^0\int_{x^0-\alpha}^{x^0+\frac{\tau}{2}}\drm x^1\int_0^{\sqrt{\left(x^0+\frac{\tau}{2}\right)^2-(x^1)^2}}\drm R\int_{\mathbb{S}_{d-3}}\drm\Omega_{d-3}\,R^{d-3}\\
	&+\int_0^{\frac{1}{2}\left(\frac{\tau}{2}+\alpha\right)}\drm x^0\int_{x^0-\alpha}^{\frac{\tau}{2}-x^0}\drm x^1\int_0^{\sqrt{\left(x^0-\frac{\tau}{2}\right)^2-(x^1)^2}}\drm R\int_{\mathbb{S}_{d-3}}\drm\Omega_{d-3}\,R^{d-3}.
\end{aligned}  
\end{equation}

These integrals are easily evaluated fixing the sprinkling spacetime dimension. For $d=4$ one gets 

\begin{equation}
	V_{+,f}^{(4,\alpha)}=\frac{\pi}{48}\tau(\tau-\alpha)(\tau+2\alpha)^2, 
\end{equation} 

which is consistent with the known formula for the volume of the whole diamond if $\alpha=0$ or $\alpha\rightarrow\tau/2$. In order to be useful, the volume must be expressed in a Lorentz invariant way.
The parameter $\alpha$ is not satisfactory, as it is proper to the reference frame in which the diamond is symmetric. Considering null coordinates $v$ and $u$ defined centred on $q$, one can see that the horizon intersects the lower diamond boundary at a point $\lambda$ away from $q$ in the $u$ direction. This is useful to define the relation

\begin{equation}
	\lambda=\frac{1}{\sqrt{2}}\left(\frac{\tau}{2}+\alpha\right)\qquad\Rightarrow\qquad\frac{\alpha}{\tau}=\sqrt{2}\frac{\lambda}{\tau}-\frac{1}{2}.
\end{equation} 

Identifying the quantity $\frac{\tau}{\sqrt{2}}=u_p$ , $\tau=\sqrt{2u_pv_p}$ and calling $c$ the ratio $c=\lambda/u_p$, one gets the form 

\begin{eqnarray}
	V_{+,f}^{(4,c)}=\frac{\pi}{4}\tau^4\left(\frac{1}{2}c^2-\frac{1}{3}c^3\right) = \pi (v_pu_p)^2\left(\frac{1}{2}\left(\frac{\lambda}{u_p}\right)^2-\frac{1}{3}\left(\frac{\lambda}{u_p}\right)^3\right).
\end{eqnarray}

We are now ready to estimate $I_n^{(4,\mathrm{flat})}$, which takes the form 

\begin{align}
	I_n^{(4,\mathrm{flat})}(l,\lambda)=&\frac{l^{-(4n+2)}}{n!}\int_\lambda^\infty\drm u\int_0^\infty\drm v\,\pi^n(vu)^{2n}\left(\frac{1}{2}\left(\frac{\lambda}{u_p}\right)^2-\frac{1}{3}\left(\frac{\lambda}{u_p}\right)^3\right)^ne^{-l^{-4} \frac{\pi}{6}\left(vu\right)^{2}},\nonumber\\
	=& \frac{\cancel{l^{-(4n+2)}}}{n!}\int_\lambda^\infty\drm u\,\cancel{l^{4n+2}}\lambda^{2n}\sqrt{\frac{3}{2\pi}}\Gamma\left[\frac{1}{2}+n\right]u^{-1-3n}\left(3u-2\lambda\right)^n,\nonumber\\
	=& \Gamma\left[\frac{1}{2}+n\right]\frac{1}{2n!}\sqrt{\frac{3}{2\pi}}\left(\frac{2^{1-2n}27^n\Gamma(2n)}{\prod_{j=1}^{2n}(-n-j)} -\frac{ \prescript{}{2}{F}_1\left(1,1-2n,1-3n;\frac{3}{2}\right) }{3n}\right)=a_n^{(4)}.
\end{align}

We obtain a result independent on the discreteness scale and on the affine distance between the horizon and the point $q$. Claim \eqref{claim_M} follows.

\section{Curved spacetimes}
 \label{curved}
 
 The results of the previous sections suggest that the expected number of horizon molecules is indeed proportional to the area of the intersection between the horizon and $\s$, when evaluated on a causal set approximating a Minkowski spacetime and once the continuum limit is taken. In generic curved spacetimes without any symmetries, however, it's possible that the limit might receive some geometric contributions that is characterized by the scale $\lambda$ introduced in the problem. Recall that in the two previous examples, $\lambda$ is the affine distance between the fold, or the point $q$, and the horizon. Since $\lambda$ itself can be arbitrarily rescaled, here we expect the contributions to have the invariant form combining $\lambda$ and curvature tensors on the null hypersurface.  For example, suppose the null congruence along $\Sigma$ intersecting the horizon has zero expansion and shear at the intersection $\J$, we ask how much does the cross section area expand/shrink at affine distance $\lambda$ away from the $\J$. Suppose $\lambda$ is small as compared to the intrinsic curvature scale at $\J$, the Raychaudhuri equation predicts the area increment to be approximately
 \begin{equation}
     \frac{\Delta A}{A} = - \int_0^\lambda \int_0^\lambda  \mathrm{Ric}(l,l)\,\dd u\,\dd u'\,.
 \end{equation}

 This dimensionless geometric quantity is of course invariant under reparameterization of the affine parameter, for that we also need to rescale the null generators $l$ in accordance with the $\lambda$ rescaling. For the problem at hand, the horizon molecule count can in principle depend on other geometric quantities that characterized by $\lambda$, the intrinsic and extrinsic curvature data on the null hypersurface $\Sigma$. On the other hand, one might ask why the scale $\lambda$ isn't apparent in the two examples evaluated in Minkowski spacetime. In the folded null plane case, there is no intrinsic or extrinsic curvatures on $\Sigma$, so there is no curvature tensor to pair with $\lambda$. In the downward light-cone case, however, it does have non-zero extrinsic curvature. But the extrinsic curvature is directly given by $1/\lambda$ so it cancels with $\lambda$ when we take the dimensionless combination. We therefore expect a possibly different behaviour of the expected number of horizon molecules in not only in generic curved spacetimes, but also whenever the null hypersurface itself has some non-trivial extrinsic curvatures. In this section, we shall only focus on the effect of the intrinsic curvature in curved spacetimes.\\
 
\subsection{Folded null planes}
Let us now consider a generic $d$-dimensional globally hyperbolic spacetime $(\mathcal{M},g)$ with a  subregion $\mathcal{H}$ defining a causal horizon. Let $\Sigma$ be a null hypersurface transverse to the horizon and intersecting it in a co-dimension $2$ surface $\mathcal{J}=\s\cap\h$. One can define coordinates adapted to this setting as follows \cite{GNC}: choose first a set of generic coordinates $y^{\alpha}$ on an open subset $\tilde{\mathcal{J}}$ of $\J$. On a neighbourhood of  $\tilde{\mathcal{J}}$ in $\mathcal{H}$ let $k^a$ be a smooth non-vanishing vector field such that the integral curves of $k^a$ are the null geodesic generators of $\h$. Assume $k^a$ is future directed. On an open neighbourhood $\tilde{\h}$ of $\tilde{\J}$ one can take as coordinates the $(d-1)$-tuple $(v,y^{\alpha})$, with $v$ a parameter running on the integral curves of $k^a$. At each point $p\in\tilde{\h}$ one can find a unique null vector field $l^a$ normal to the spatial direction of $\h$ and such that $k^al_a=-1$, so that it is also future directed. We can build coordinates $(v,u,y^{\alpha}) $ on an open neighbourhood $\mathcal{N}$ of $\tilde{\h}$ with $u$ the affine parameter along null geodesics generated by $l^a$ at the point $(v,y^{\alpha})$ on $\tilde{\h}$. These coordinates are known in the literature as Gaussian Null coordinates (GNC). We define now a null hypersurface $\B$ as the collection of the integral lines of the vector field $k^a_{\lambda}$, where $k^a_{\lambda}$ is the vector field obtained by parallel transporting $k^a$ along the geodesics defining the surface $\s$ of a parameter $s=\lambda$. \\
 
 Consider again a causal set $(\mathcal{C},\prec) $ sprinkled into $\mathcal{M}$ with a Poisson point process of density $\rho=l^{-d}$. In the tubular neighbourhood $\mathcal{N}\supset\J$ in which GNC are constructed define the region 
 
 \begin{equation}
 	\mathcal{R}_{\Lambda}:=\{p\in I^+(\J)\cap\mathcal{N}\,|\,0<v(p)<-\Lambda,\, 0<u(p)<-\Lambda\}, 
 \end{equation} 
 
 where $\Lambda$ sets an intermediate scale between the discreteness length of $\mathcal{C}$ and the geometric length scale of the problem, i.e. $l\ll \Lambda\ll L_G$. In order for the entropy to be a quantity local to the intersection $\J$, one hopes to show equation \eqref{entropy_def_b} to be dominated by the integral on $\mathcal{R}_{\Lambda}$ 
 
 \begin{equation}
 	\rho^{\frac{2-d}{d}}\big\langle\mathbf{H}_n\big\rangle=\rho^{\frac{2-d}{d}+1}\int_{\mathcal{R}_{\Lambda}}\,\drm V_p \,\frac{(\rho V_+(p;\lambda))^n}{n!}e^{-\rho V(p;\lambda)}+\dots,
 	\label{local_a}
 \end{equation}
 
 with $\dots$ denoting terms decaying exponentially fast in the limit of $l\rightarrow0$. As already discussed in the flat spacetime case, we believe this is not true in general in the case of $\s$ being a null hypersurface. The argument goes as follows: when $v(p)\leq-\Lambda$ and for all values of $u(p)$, the volume $V(p)$ will be large enough so that the exponential term in \eqref{entropy_def} effectively suppresses the contributions coming from this region in the continuum limit. In this case, one can suppose the set of values of $V(p)$ foliates the integration domain and the bounds of \cite{Dowker}, section  IV.B, apply.\\
 
 On the other hand, when $u(p)\leq-\Lambda$ this argument fails, as one can pick the point $p$ to be close to the axis $v(p)=0$ so that the volume $V(p)$ is close to zero for values of $u$ arbitrarily far in the past of $\J$. Thus, the integral would in principle receive contributions coming from far away along the past light-cone of the intersection hypersurface. These contributions are not guaranteed to be convergent, as the coordinate $u$ can run up to past infinity. In addition, they could also be ill defined, as the assumption for the curvature to be well behaved on a neighbourhood of the origin fails far away from it and caustics could in general appear along the light-cone. It is therefore difficult to give a formal treatment of the integral behaviour in this region. Therefore, we believe that the presence of information coming from far away in the past of the origin, even in the continuum limit, is a first indication the proposal of counting horizon molecules with a null hypersurface $\s$ is a flawed way to define entropy for a causal set. \\
 
 One could however ask if the proposal of Barton et al together with a null hypersurface is still viable after considering $\Lambda$ as a IR cutoff in the $u$ direction, in order to exclude contributions from the far past of the intersection in the final count of the entropy. In the next subsection, where we consider a finite segment of null hypersurface attached with spacelike tails, such IR cutoff is automatically imposed by locality arguments as in Barton et al. We will therefore for the moment put it by hand and restrict our attention to the $\mathcal{R}_\Lambda$ region. We shall explicitly write the (\ref{local_a}) in GNC, always after inverting the time direction so to integrate $p$ over the future of $\J$,
\begin{equation}\label{local_b}
\begin{aligned}
\rho^{\frac{2-d}{d}}\big\langle\mathbf{H}_n\big\rangle=&\frac{\rho^{\frac{2-d}{d}+1+n}}{n!}\int_{\J}\,\drm^{d-2} y\,\int_0^\Lambda\,\drm v\,\int_0^{\Lambda}\drm u\, \sqrt{-g(v,u,y)}\,\big(\rho V_+(v,u,y;\lambda)\big)^n e^{-\rho V(v,u,y;\lambda)}+\dots\\
=&\int_{\J}\,\drm^{d-2} y\,\sqrt{\sigma(y)}\,I_n^{(d)}(y;l,\Lambda, \lambda)+\dots,
\end{aligned}
\end{equation}
where we made explicit the dependence of the volumes on the parameter $\lambda$ and we defined $\sigma(y)$ the induced metric on $\J$. This defines the function for any point $y\in\J$,

\begin{equation}
I_n^{(d)}(y;l,\Lambda, \lambda):=\frac{l^{-(dn+2)}}{n!}\int_0^\Lambda\,\drm v\,\int_0^{\Lambda}\drm u\, \sqrt{-\frac{g(v,u,y)}{\sigma(y)}}\,\big( V_+(v,u,y;\lambda)\big)^n e^{-\rho V(v,u,y;\lambda)}.
\label{local_c}
\end{equation}

Here we can expand the above function in $l$ up to the $\mathcal{O}(l)$ order. As compared to the previous calculations in the flat case where $I_n^{(d)}$ essentially localises to $\J$, we expect here deviations due to the non-vanishing curvature tensors.  Since $I_n^{(d)}$ is dimensionless,  we need to contract any curvature tensors with objects with length dimensions. There are three independent scales in this problem: $l$, $\lambda$ and $\Lambda$. They correspond to the discreteness scale, the distance between the fold and the horizon and the cutoff scale.  By dimensional analysis and locality arguments, $I_n^{(d)}(y;l,\Lambda, \lambda)$ admits a small $l$ expansion of the form 

\begin{equation}
I_n^{(d)}(y;l,\Lambda, \lambda)=a_n^{(d)}+\sum_jc_{n,j}^{(d)}\mathcal{F}_j(y,\lambda,\Lambda)+l\sum_ib_{n,i}^{(d)}\,\mathcal{G}_i(y,\lambda,\Lambda)+\mathcal{O}(l^2),
\label{I_expansion}
\end{equation}

where $a_n^{(d)}$, $b_{n,i}^{(d)}$ and $c_{n,j}^{(d)}$ are constants dependent on $d$ and $n$. The set $\left\{\mathcal{G}_i(y,\lambda,\Lambda)\right\}$ is the set of mutually independent geometric scalars of length dimension $L^{-1}$ evaluated on the geodesic segment $\gamma_q(s)$. Likewise, the set $\left\{\mathcal{F}_j(y,\lambda,\Lambda)\right\}$ is the set of mutually independent dimensionless geometric scalars  evaluated on the geodesic segment $\gamma_q(s)$. These scalars can be obtained by from curvature tenors contracting with objects carrying scale $\lambda$ or $\Lambda$. Equation \eqref{I_expansion} implies

\begin{equation}
\lim_{l\rightarrow0}I_n^{(d)}(y;l,\Lambda, \lambda)=a_n^{(d)}+\sum_j c_{n,j}^{(d)}\mathcal{F}_j(y,\lambda,\Lambda).
\label{I_limit}
\end{equation}

It is worth noting the difference between our analysis and  \cite{Dowker} in the spacelike case. The case thereby discussed sees the volumes $V$ and $V_+$ tend to the point $y$ in the continuum limit and the integrand of $I_n^{(d)}$ is non-negligible in a neighbourhood of this point. The small $l$ expansion is therefore dependent only on the geometric scalars evaluated at $y$. There are no other scales than $l$ which can pair with the curvature tensors, so their $I_n^{(d)}$ does not have any such  $\left\{\mathcal{F}_j(y,\lambda,\Lambda)\right\}$ contributions. Their corresponding $I_n^{(d)}$ reads 

\begin{equation}
    I_n^{(d)}(y;l,\tau)=a_n^{(d)}+l\sum_ib_{n,i}^{(d)}\,\mathcal{G}_i(y)+\mathcal{O}(l^2)
\end{equation}
where $\tau$ is a middle scale corresponding to our $\Lambda$ and the curvature contributions are all localized to $y\in\J$.\\

In our null setup,  however, the small $l$ expansion must include geometric information evaluated on the geodesic segment from the horizon to the hypersurface $\B$, which carries with it additional scales $\lambda$ and $\Lambda$. Since such contributions $\left\{\mathcal{F}_j(y,\lambda,\Lambda)\right\}$ are not forbidden on dimensional grounds, we expect generically part of these corrections to be present also in the continuum limit, which is summarised by equation \eqref{I_limit}. One can verify this by considering mild curvature perturbations close to $\J$. These perturbations will enter the volume expressions $V,V_+$ in (\ref{local_c}), and they generally do not cancel out. In the next subsection, we confirm this limiting behaviour in the downward light-cone example with explicit calculations.

\subsection{Downward Light-cone}

We consider the same setup as in section \ref{LC_f}. Consider $\Sigma$ as the past-pointing light-cone of the point $q$, which is of affine distance $\lambda$ away from $\J$. The codimension-$2$ region $\J$ is defined as $\J:=\partial I^-(q)\cap\h$. Again, the volumes of interest are

\begin{equation}
\begin{aligned}
V_+(p):=&\mathrm{vol}\big( I_{[p,q]}\cap   I^+(\h) \big),\\
V(p):=&\mathrm{vol}\big( I_{[p,q]}).
\end{aligned} 
\end{equation}

\begin{figure}[h]
	\centering 
	\begin{tikzpicture}[scale=1]
	
	\fill[gray!20] (0,0) -- (2,-2) -- (2.5,-1.5) -- (.5,.5) -- cycle;
 	\draw[-] (-1.5,1.5)--(2.5,-2.5);
	\draw[-] (0,1)-- (2.5,-1.5);
	\draw[-] (-1,-1)-- (1.5,1.5);
	\draw[-] (0,1)-- (-.5,.5);
	\draw[-] (2,-2)--(2.5,-1.5);

	\node at (1.4,1.8) {$\mathcal{H}$};
	\node at (-.5,0) {$\mathcal{J}$};
	
	\draw[->,line width=0.3mm] (-0.5,-0.5)--(.8,.8);
	\node at (1,.7) {$v$};
	\draw[->,line width=0.3mm] (.5,-.5)--(-.8,.8);
	\node at (-1,.7) {$u$};
	
	\node at (0,1)[circle,fill,inner sep=1pt]{};
	\node at (0,1.2) {$p$};
	\node at (0,0)[circle,fill,inner sep=1pt]{};
	\node at (0.4,0) {$y$};
	\node at (2,-2)[circle,fill,inner sep=1pt]{};
	\node at (2.3,-2) {$q$};
	
	\draw[line width=0.25mm,decoration={brace,mirror,raise=5pt},decorate]
	(0,0) -- (2,-2);
	
	\node at (.65,-1.3) {$\lambda$};
	\node at (.-1.8,1.2) {$\gamma(s)$};
	
	\end{tikzpicture}
	\caption{Time-reversed coordinate system in the skinny diamond setup. The volume $V_+$ is shaded.}
	\label{Hm-NFNC}
\end{figure}
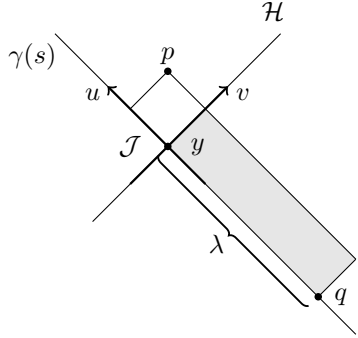

There will be a unique null geodesic $\gamma$ passing through $q$, being transverse to the generators of the future light-cone of $p$ and crossing the horizon at a point $y$. Once again we will reverse the temporal direction to perform the computations, so that $p\in I^+(\h)$ and $q\in I^-(\h)$. We will assume this geodesic to be affinely parametrised so that $\gamma(0)=y$, $\gamma(-\lambda)=q$. In the continuum limit we expect the point $p$ to shrink towards $\gamma$, and the volumes $V$ and $V_+$ shall tend to a skinny causal interval. Thus, we can set up a Null Fermi Normal Coordinates system $(v,u,y^\alpha)$ centred on this geodesic and with origin on $\J$, where $y^\alpha$ are carried from the local coordinates of $y$ in $\J$.  The horizon is given by the surface $u=0$. A sketch of the coordinate system is given in Figure \ref{Hm-NFNC}.\\

As in the previous section, the horizon molecules expectation value can be written as \eqref{local_b} in a tubular neighbourhood of $\J$ containing the point $q$ and controlled by the parameter $\Lambda$. Thus we are once again left with the function $I_n^{(d)}(y;l,\Lambda,\lambda)$ defined as in equation \eqref{local_c}. For the sake of simplicity we consider a spacetime dimension $d=4$. We assume $\lambda,\Lambda$ to be small relative to the local curvature scales such that we can obtain perturbative volume expansions for $V$ and $V_+$ in $u,v$ using the geometric data on $\gamma$. These were computed in the Appendix. Calling $(v,u)$ the null coordinates of point $p$ and fix a point $y\in\J$, the volume expansion reads as (cf. (\ref{volumeformula}))

\begin{align}
V^{(4)}(u,v,y;\lambda)&=\frac{\pi\tau^4}{24}\Bigg(1+12\int_{-\lambda}^u\drm u'\frac{(u-u')^2}{(u+\lambda)^3(u'+\lambda)}\int_{-\lambda}^{u'}\drm x\int_{-\lambda}^x \drm x'\,(x'+\lambda)R_{u'u'}(x'+\lambda)\Bigg)+\mathcal{O}((u+\lambda)^3,v^3),\nonumber\\
&=V^{(4)}_f\Bigg(1+F(\lambda,u)\Bigg)+\mathcal{O}((u+\lambda)^3,v^3)
\label{V_skinny}
\end{align}
where we define $F(\lambda,u)$ to denote the integral. Analogously, defining $c$ the ratio $c=\lambda/(u+\lambda)$, the truncated diamond volume becomes

{\small
\begin{align}
V_+^{(4)}(u,v,y;\lambda)&=\frac{\pi\tau^4}{4}\left(\frac{c^2}{2}-\frac{c^3}{3}\right)+\frac{\pi\tau^4}{2}\int_{-\lambda}^0\drm u'\frac{(u'+\lambda)^2(u-u')^2}{(u+\lambda)^6}\int_{-\lambda}^{u'}\drm x\int_{-\lambda}^x \drm x'\,(x'+\lambda)R_{u'u'}(x'+\lambda)+\mathcal{O}((u+\lambda)^3,v^3),\nonumber\\
&=V^{(4)}_{+,f}\Bigg(1+2\left(\frac{c^2}{2}-\frac{c^3}{3}\right)^{-1}\int_{-\lambda}^0\drm u'\frac{(u'+\lambda)^2(u-u')^2}{(u+\lambda)^6}\int_{-\lambda}^{u'}\drm x\int_{-\lambda}^x \drm x'\,(x'+\lambda)R_{u'u'}(x'+\lambda)\Bigg)+\mathcal{O}((u+\lambda)^3,v^3),\nonumber\\
&=V^{(4)}_{+,f}\Bigg(1+\tilde{F}(\lambda,u)\Bigg)+\mathcal{O}((u+\lambda)^3,v^3)
\label{V+_skinny}
\end{align}
}%
where we define $\tilde{F}(\lambda,u)$ to denote the integral. \\

Our goal is to show that the continuum limit of $I_n^{(d)}$ contains non-vanishing curvature terms. We can now insert those expansions in equation \eqref{local_c} and ignore henceforth the $\mathcal{O}(v^3,(u+\lambda)^3)$ tails as we assume $\lambda, \Lambda$ are much smaller than the curvature scales. 

{\small
\begin{align}
I_n^{(d)}(y;l,\Lambda, \lambda)=&\frac{l^{-(4n+2)}}{n!}\int_0^\Lambda\,\drm v\,\int_0^{\Lambda}\drm u\,e^{-l^{-4} V^{(4)}_f(u,v,y;\lambda)}\left(V^{(4)}_{+,f}(u,v,y;\lambda)\right)^n\Bigg(1-l^{-4}V^{(4)}_fF(\lambda, u)+n\tilde{F}(\lambda, u)\Bigg),\nonumber\\
=&\frac{l^{-(4n+2)}}{n!}\int_0^\Lambda\,\drm v\,\int_0^{\Lambda}\drm u\,e^{-l^{-4} V^{(4)}_f(u,v,y;\lambda)}\left(V^{(4)}_{+,f}(u,v,y;\lambda)\right)^n\nonumber\\
&-\frac{l^{-(4n+2)}}{n!}\int_0^\Lambda\,\drm v\,\int_0^{\Lambda}\drm u\,e^{-l^{-4} V^{(4)}_f(u,v,y;\lambda)}\left(V^{(4)}_{+,f}(u,v,y;\lambda)\right)^nl^{-4}V^{(4)}_fF(\lambda, u)\nonumber\\
&+\frac{l^{-(4n+2)}}{n!}\int_0^\Lambda\,\drm v\,\int_0^{\Lambda}\drm u\,e^{-l^{-4} V^{(4)}_f(u,v,y;\lambda)}\left(V^{(4)}_{+,f}(u,v,y;\lambda)\right)^n
n\tilde{F}(\lambda, u).
\label{explicit_skinny}
\end{align}
}%

The first term of equation \eqref{explicit_skinny} is the all-flat term computed in section \ref{LC_f}. Assuming the integrals in the second and third lines are suppressed exponentially fast for $v>\Lambda$, we can write

\begin{align}
&-\frac{l^{-(4n+2)}}{n!}\int_0^\Lambda\,\drm v\,\int_0^{\Lambda}\drm u\,e^{-l^{-4} V^{(4)}_f(u,v,\lambda)}\left(V^{(4)}_{+,f}(u,v,y;\lambda)\right)^nl^{-4}V^{(4)}_f(u,v,y;\lambda)F(\lambda, u),\nonumber\\
&=-\frac{\sqrt{\frac{3}{2\pi}}\Gamma\left(\frac{3}{2}+n\right)}{\Gamma\left(n+1\right)}\int_0^\Lambda\drm u\,\frac{F(\lambda,u)}{u+\lambda}\left(3\left(\frac{\lambda}{u+\lambda}\right)^2-2\left(\frac{\lambda}{u+\lambda}\right)^3\right)^n+...,
\label{LC_nf_V}
\end{align}

where $...$ denote terms decaying exponentially fast in the limit $l\rightarrow0$. Analogously, the third term becomes

\begin{align}
&\frac{l^{-(4n+2)}}{(n-1)!}\int_0^\Lambda\,\drm v\,\int_0^{\Lambda}\drm u\,e^{-l^{-4} V^{(4)}_f(u,v,y;\lambda)}\left(V^{(4)}_{+,f}(u,v,y;\lambda)\right)^n\tilde{F}(\lambda, u),\nonumber\\
&=\frac{\sqrt{\frac{3}{2\pi}}\Gamma\left(\frac{1}{2}+n\right)}{\Gamma\left(n\right)}\int_0^\Lambda\drm u\,\frac{\tilde{F}(\lambda,u)}{u+\lambda}\left(3\left(\frac{\lambda}{u+\lambda}\right)^2-2\left(\frac{\lambda}{u+\lambda}\right)^3\right)^n+...
\label{LC_nf_V+}.
\end{align}

We can now evaluate $F(\lambda,u)$ and $\tilde{F}(\lambda,u)$. For simplicity we assume that $R_{u'u'}(y)=\mathcal{R}(y)$ is constant over $u$, which is good enough to support our claim. Thus, 

\begin{align}
    F(\lambda, u)&=\frac{\mathcal{R}(y)}{15}(u+\lambda)^2, & \tilde{F}(\lambda, u)&=\frac{\mathcal{R}(y)}{90}\left(\frac{1}{2}\left(\frac{\lambda}{u+\lambda}\right)^2-\frac{1}{3}\left(\frac{\lambda}{u+\lambda}\right)^3\right)^{-1}\frac{\lambda^3 (\lambda^2 + 5\lambda u + 10 u^2)}{(\lambda+ u)^3}.
    \label{F_Ftilde}
\end{align}

Inserting now \eqref{F_Ftilde} into \eqref{LC_nf_V} and \eqref{LC_nf_V+} we get 

\begin{align}
&-\frac{\sqrt{\frac{3}{2\pi}}\Gamma\left(\frac{3}{2}+n\right)}{\Gamma\left(n+1\right)}\int_0^\Lambda\drm u\,\frac{F(\lambda,u)}{u+\lambda}\left(3\left(\frac{\lambda}{u+\lambda}\right)^2-2\left(\frac{\lambda}{u+\lambda}\right)^3\right)^n,\nonumber\\
=&-\frac{\sqrt{\frac{3}{2\pi}}\Gamma\left(\frac{3}{2}+n\right)\mathcal{R}(y)}{15\Gamma\left(n+1\right)} \int_0^\Lambda\drm u\,(u+\lambda)\left(3\left(\frac{\lambda}{u+\lambda}\right)^2-2\left(\frac{\lambda}{u+\lambda}\right)^3\right)^n\sim
    \begin{cases}
      \mathcal{R}(y)\lambda^2 \log(\lambda+\Lambda) & \text{if } n=1\\
      \mathcal{R}(y)\lambda^2 f(\lambda,\Lambda) & \text{if } n>1,\\
      \end{cases}
\label{F_example_final}
\end{align}

with $f(\lambda,\Lambda)\sim\mathcal{O}(1)$ if $\Lambda\gg\lambda$; and 

\begin{align}
&\frac{\sqrt{\frac{3}{2\pi}}\Gamma\left(\frac{1}{2}+n\right)}{\Gamma\left(n\right)}\int_0^\Lambda\drm u\,\frac{\tilde{F}(\lambda,u)}{u+\lambda}\left(3\left(\frac{\lambda}{u+\lambda}\right)^2-2\left(\frac{\lambda}{u+\lambda}\right)^3\right)^n,\nonumber\\
=&\frac{\sqrt{\frac{3}{2\pi}}\Gamma\left(\frac{1}{2}+n\right)\mathcal{R}(y)}{15\Gamma\left(n\right)}\int_0^\Lambda\drm u\, \frac{\lambda^3(\lambda^2 + 5 \lambda u + 10 u^2)}{(\lambda + u)^4}\left(\frac{\lambda^2(\lambda+3u)}{(\lambda+u)^2}\right)^{n-1}\sim \mathcal{R}(y)[\lambda^2+\tilde{f}(\lambda,\Lambda)],
\end{align}

with $\tilde{f}(\lambda,\Lambda)\rightarrow0$ if $\Lambda\gg\lambda$. Thus, the horizon molecules expectation value admits a continuum limit of the form \eqref{I_limit}, i.e.

\begin{equation}
\lim_{l\rightarrow0}I_n^{(d)}(y;l,\Lambda, \lambda)= a_n^{(d)}+\mathcal{R}(y)c^{(d)}_n(\lambda,\Lambda)+\cdots
\end{equation}

where the $\cdots$ contains other curvature terms that are higher order in the volume expansions, e.g. $\sim R^2\lambda^4$. We see that the limit is not local to the intersection $\J$.  We thus conclude that this horizon molecule definition does not yield a well behaved area law for the entropy when evaluated on a null hypersurface crossing a causal horizon.

\subsection{Hypersurfaces of mixed signature}\label{sec:mixed}
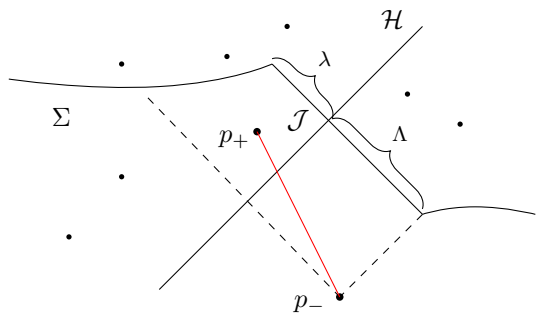
\begin{figure}[h]
\centering 
\begin{tikzpicture}[scale=1]

\draw [-] plot [smooth, tension=1] coordinates { (-2.5,.8) (-.5,0.7) (1,1)};
\draw[-] (1,1) -- (3,-1);
\draw [decorate,decoration={brace,amplitude=7pt},xshift=0pt,yshift=2pt]
(1,1) -- (1.8,0.2) node [black,midway,yshift=0.4cm,xshift=0.3cm] 
{\footnotesize $\lambda$};
\draw [decorate,decoration={brace,amplitude=7pt},xshift=0pt,yshift=2pt]
(1.8,0.2) -- (3,-1) node [black,midway,yshift=0.4cm,xshift=0.3cm] 
{\footnotesize $\Lambda$};
\draw [-] plot [smooth, tension=1] coordinates { (3,-1) (3.7,-.9) (4.5,-1)};
\draw[-] (-.5,-2) -- (3,1.5);

\node at (1.9,-2.1)[circle,fill,inner sep=1pt]{};
\node at (.8,.1)[circle,fill,inner sep=1pt]{};

\draw[-, dashed] (3,-1) -- (1.9,-2.1);
\draw[-, dashed] (-0.65,0.55) -- (1.9,-2.1);

\draw[-,red] (1.9,-2.1) -- (.8,0.1);

\node at (1.5,-2.2) {$p_-$};
\node at (.5,0) {$p_+$};
\node at (-1.8,.3) {$\Sigma$};
\node at (2.6,1.6) {$\mathcal{H}$};
\node at (1.35,.25) {$\mathcal{J}$};

\node at (.4,1.1)[circle,fill,inner sep=.7pt]{};
\node at (1.2,1.5)[circle,fill,inner sep=.7pt]{};
\node at (2.8,.6)[circle,fill,inner sep=.7pt]{};
\node at (-1,-.5)[circle,fill,inner sep=.7pt]{};
\node at (-1.7,-1.3)[circle,fill,inner sep=.7pt]{};
\node at (3.5,.2)[circle,fill,inner sep=.7pt]{};
\node at (-1,1)[circle,fill,inner sep=.7pt]{};

\end{tikzpicture}
\caption{Mixed signature hypersurface $\Sigma$ which is everywhere spacelike except for a null segment crossing the horizon. The contribution from the region beyond the dashed line is exponentially suppressed. }
\label{Mixed_signature}
\end{figure}

So far we have studied the particular example of downward light-cone and shown that horizon molecule count in the continuum limit is not proportional to the horizon area, but rather it is also influenced by the curvature data on the null surface. It can be traced down to the fact that the dominant region of contribution localizes to the entire neighbourhood of bounded null segment, rather than $\J$ as in the spacelike case. As we argued from the dimensional grounds, we believe this dependence is generic whenever we have a null hypersurface crossing the horizon. In particular, this can even be a piece of null segment that is part of an elsewhere spacelike hypersurface as illustrated in figure \ref{Mixed_signature}. \\

Consider such a hypersurface of mixed signature, whose null segments spans a parameter distance $\lambda+\Lambda$. The advantage of this setup as opposed to the infinitely past-extending null hypersurface is that the spacelike sections provide natural IR cutoffs as we alluded to earlier. One can apply the locality argument of Barton et al up to the point where the local neighbourhood to $\J$ crosses the spacelike sections and contains mostly the null segments. (See the shaded region in Figure \ref{Mixed_signature}.) This defines for us a cutoff scale $\Lambda$ beyond which the contribution is exponentially suppressed. Then we just need to compute (\ref{local_c}) and it's clear from above arguments that generically this integral is not going to localize to $\J$ as we approach the continuum limit. Note that the hypersurface signature at the intersection $\J$ needs not be everywhere null for this argument to hold. The null segment can also have finite extent in the transverse direction, which is enough to distort the area law behaviour. 

\section{Entropy from SMI}
\label{SMI}

\begin{figure}[h]
	\centering 
	\begin{tikzpicture}[scale=1]
		\draw [-] plot [smooth, tension=1] coordinates { (-2.5,-.2) (0,0.1) (3,-.2) (4.5,.1)};
		\draw[-] (-1,-2.5) -- (2.3,1.5);

		\node at (-1.8,.3) {$\Sigma$};
		\node at (2.6,1.6) {$\mathcal{H}$};
		\node at (-1,-.8) {$X$};
		\node at (1.2,-1) {$Y$};
		
		\node at (.4,1.1)[circle,fill,inner sep=.7pt]{};
		\node at (1.2,1.5)[circle,fill,inner sep=.7pt]{};
		\node at (.2,0.6)[circle,fill,inner sep=.7pt]{};
		\node at (2.8,.6)[circle,fill,inner sep=.7pt]{};
		\node at (-.5,-.5)[circle,fill,inner sep=.7pt]{};
		\node at (-1.7,-1.3)[circle,fill,inner sep=.7pt]{};
		\node at (1.9,-.8)[circle,fill,inner sep=.7pt]{};
		\node at (2.5,-1.3)[circle,fill,inner sep=.7pt]{};
		\node at (3.5,.2)[circle,fill,inner sep=.7pt]{};
		\node at (-1,1)[circle,fill,inner sep=.7pt]{};
		
	\end{tikzpicture}
	\caption{Partition of a spacetime and of a causal set by a causal horizon $\h$.}
	\label{XYpartition}
\end{figure}
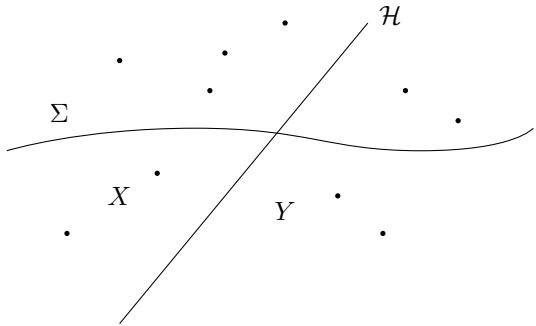

In the previous sections, we discussed the issues of applying the horizon molecules program to null $\s$ hypersurfaces. To still be able to properly define a horizon entropy in this situation, we now focus on the \textit{spacetime mutual information} (SMI) proposal. Take a spacetime $(\M,g)$ approximated by a causal set $(\C,\prec)$. A causal horizon $\h$ can be given in the causal set, mapping a future-inextensible timelike curve in $\M$ to a future-infinite chain of causal set elements. Considering then a non-timelike hypersurface $\s$ crossing the horizon, which can also be defined in similar ways in a causal set, one is left with two regions $X$ and $Y$ laying in the past of $\s$ and separated by $\h$ (Figure \ref{XYpartition}). Defining $Y$ as the region at the past of the horizon, we see that it will evolve independently of $X$ as far as causality is preserved. \\

In \cite{Benincasa}, this latest fact together with the properties of the Causal Set action led to define a quantity that can be a candidate for a horizon entropy. Recall the definition of the Benincasa-Dowker action for causal sets in dimension $4$ \cite{benincasa2010scalar,dowker2013causal},

\begin{equation}
	S^{(4)}(C)=\frac{4}{\sqrt{6}l^2}\Big[N-N_0+9N_1-16N_2+8N_3\Big],
	\label{BD_act}
\end{equation}

where $N$ is the cardinality of the causal set, $l$ is a fundamental length, $N_m$ the number of $m$-inclusive intervals in $\C$. If one partitions the causal set $\C$ into two subsets $X$ and $Y$ so that $\C=X\cup Y$ and $X\cap Y=\emptyset$ the action, noted from now on by $S$ for shortness, will not be local and additive, i.e.

\begin{equation}
	S[\mathcal{C}]\neq S[X]+S[Y].
\end{equation}

Taking the cue from thermodynamics and information theory, one can therefore define the Spacetime Mutual Information (SMI) between two regions $X$ and $Y$ as 

\begin{equation}
	\mathcal{I}^{(d)}_\s[X,Y]:=\left(\frac{l_p}{l}\right)^{d-2}\left(S^{(d)}_{BD}[X]+S^{(d)}_{BD}[Y]-S^{(d)}_{BD}[\C]\right).
	\label{SMI_b}
\end{equation}

In \cite{Benincasa} is conjectured that the expectation value of the SMI tends, in the continuum limit, to the volume of the intersection between the horizon and the hypersurface $\s$, i.e. $\J:=\h\cap\s$, times a dimension-dependent constant, 

\begin{equation}
	\lim_{l\rightarrow 0}\left\langle \mathcal{I}^{(d)}_\s[X,Y]\right\rangle= b_d\frac{\mathrm{vol}(\J)}{l_p^{d-2}}.
	\label{conjecture}
\end{equation}

This would then suggest the SMI gives the continuum area law for the horizon entropy, at least from a kinematical point of view. The evidence to support this conjecture has been collected from numerical simulations and it seems to be consistent with the expectations. We shall thereby consider a simple setup with a causal interval in 4D Minkowski spacetime and analytically confirm (\ref{conjecture}):
\begin{equation}
	\lim_{l\rightarrow 0}\left\langle \mathcal{I}^{(4)}_\s[X,Y]\right\rangle= \frac{\mathrm{vol}(\J)}{l_p^2}.
	\label{result2}
\end{equation}
with $b_4=1$.\\

We demonstrate the calculations in Minkowski spacetime $(\mathbb{M}^4,\eta)$. We suppose the spacetime is partitioned by a Rindler causal horizon $\h$. We pick two points in $\mathbb{M}^4$, $p$ and $q$, sitting in the past and in the future of the horizon respectively. And we consider the causal diamond $I[p,q]$ between these two points. We define Null Fermi Normal Coordinates, set up as in the Appendix. From the point $p$, we shoot a null geodesic $\gamma(p,o)$ of affine length $L$. At $o$, a second null geodesic orthogonal to $\gamma(p,o)$ is shot, defining $\gamma'(o,q)$, with $q$ at an affine parameter $s$ along it. This defines the causal interval $I[p,q]$. We then define Null Fermi Normal Coordinates adapted to the geodesic $\gamma(p,o)$, $(x^+,x^-,r,\theta)$. Note the transverse spatial part is described in spherical coordinates. Thus, the metric reads

\begin{equation}
	\drm s^2=-2\drm x^+\drm x^-+\drm r^2+r^2\drm\theta^2
	\label{metric_NFNC_M}
\end{equation}

and the relevant points in our construction have coordinates $p^\mu=(0,...,0)$, $o^\mu=(L,0,...,0)$ and $q^\mu=(L,s,0,...,0)$. The horizon is given by the hypersurface $\h=\{(cL,x^-,r,\theta)\}$, with $c\in(0,1).$\\

Consider now as usual a causal set $\mathcal{C}$ sprinkled in the interval $I[p,q]$ at a density $\rho$. In this setup, the causal diamond is partitioned by the horizon into two regions $X$ and $Y$, with $X$ lying in the future of the horizon and $Y$ in its past. The computation of the spacetime mutual information between these two regions can be tackled in different ways, either looking for $N_0(X,Y|\mathcal{C})$ or by computing the action of the regions $X$ and $Y$ separately. The first approach turns out to be difficult, especially because of the reduced spherical symmetry of the problem due to the presence of the horizon as a boundary of $X$ and $Y$. Looking for the action on the two subregions is easier, especially noticing that the region $X$ has the horizon sitting only on the past boundary. The inner integral in the usual action evaluation is thus blind to the presence of the horizon and can be straightforwardly carried out considering a properly boosted causal diamond. One has 

\begin{align}
	\left\langle N_0^{(4)}(X)\right\rangle&=\rho^2\int_X\drm^4x\int_{X\cap I^+(x)}\drm^4y\,e^{-\rho V_{xy}}\nonumber\\
	&=\rho^2\sum_n^\infty\frac{\left(-\rho\frac{\pi}{24}\right)^n}{n!}\int_X\drm^4x \frac{\pi\Gamma(2n+4)}{4(n+1)\Gamma(2n+4)}\tau_{xq}^{4(n+1)}.
	\label{N0_X_4d}
\end{align}

The integration measure over $X$ becomes in NFNC

\begin{equation}
	\int_X\drm^4x=\int_{cL}^L\drm x^+\int_0^{r*}\drm r\,r\int_{x^-_{min}}^{x^-_{max}}\drm x^-\int_{S_{1}}\drm\theta,
	\label{measure_NFNC}
\end{equation}

with

\begin{align}
	x^-_{min}=\frac{r^2}{2x^+},\quad\quad  x^-_{max}=\frac{r^2}{2(x^+-L)}+s,
\end{align}

fixing the boundaries of the causal diamond and $r^*=\sqrt{2sx^+(1-x^+/L)}$ is the radius at which the upper and lower light-cones intersect. Furthermore, the proper time between $x$ and $q$ is given by 

\begin{equation}
	\tau_{xq}^2=2(L-x^+)(s-x^-)-r^2.
	\label{tau_xq_M4}
\end{equation}

Equation \eqref{N0_X_4d} therefore becomes

\begin{align}
	\left\langle N_0^{(4)}(X)\right\rangle&=\rho^2\sum_n^\infty\frac{\left(-\rho\frac{\pi}{24}\right)^n}{n!}\int_X\drm^4x \frac{\pi\Gamma(2n+4)}{4(n+1)\Gamma(2n+4)}\tau_{xq}^{4(n+1)},\nonumber\\
	&=\rho^2\sum_n^\infty\frac{\left(-\rho\frac{\pi}{24}\right)^n}{n!}\frac{\pi^2\Gamma(2n+4)}{2(n+1)\Gamma(2n+4)}\int_{cL}^L\drm x^+\int_0^{r*}\drm r\nonumber\\
	&\qquad\cdot\int_{x^-_{min}}^{x^-_{max}}\drm x^-\,r\,\big(2(L-x^+)(s-x^-)-r^2\big)^{2(n+1)},\nonumber\\
	&=\rho^2\sum_n^\infty\frac{\left(-\rho\frac{\pi}{24}\right)^n}{n!}\frac{\pi^2(1+2c(n+2))(1-c)^{2n+4}}{2^6(n+1)^2(n+2)^2(2n+1)(2n+3)^2(2n+5)}(\tau^4)^{n+2},\nonumber\\
	&=\sum_n\frac{(-N_X)^{n+2}}{n!}\frac{9\,(1+2c(n+2))}{(1+2c)^{n+2}(n+1)^2(n+2)^2(2n+1)(2n+3)^2(2n+5)}.
	\label{N0_X_4d_final}
\end{align}

In the third equality we used the identity $\tau^2=2Ls$ and in the last step we wrote the sum as a function of $N_X$, i.e. the number of causal set elements sprinkled in the region $X$, which is given by 

\begin{equation}
	N_X=\rho\,\mathrm{vol}(X)=\rho\,\frac{\pi\,\tau^4}{24}(1-c)^2(1+2c).
	\label{NX}
\end{equation}

After differentiation of equation \eqref{N0_X_4d_final} with respect to $\rho$ to obtain the number of $m-$inclusive intervals in $X$, we are ready to insert the results into the definition of the causal set action in $4$ dimensions 

\begin{equation}
	\frac{1}{\hbar}\left\langle S^{(4)}_{BD}(X)\right\rangle=\alpha_4\left(\frac{l}{l_p}\right)^2\left(N_X-\left\langle N_0^{(4)}(X)\right\rangle+9\left\langle N_1^{(4)}(X)\right\rangle-16\left\langle N_2^{(4)}(X)\right\rangle+8\left\langle N_3^{(4)}(X)\right\rangle\right).
\end{equation}

Taking the continuum limit, i.e. sending $N_X$ to infinity, one gets that the action has a leading contribution

\begin{equation}
	\frac{1}{\hbar}\left\langle S^{(4)}_{BD}(X)\right\rangle\sim\,\frac{l^2}{l_p^2}2\sqrt{6\pi\,N_X}\frac{(1+c)}{\sqrt{1+2c}}.
	\label{action_X_a}
\end{equation}

Inserting equation \eqref{NX} into \eqref{action_X_a} we can express the continuum limit as a function of the number of points sprinkled on the whole $I[p,q ]$ interval, which then reads

\begin{equation}
	\frac{1}{\hbar}\left\langle S^{(4)}_{BD}(X)\right\rangle\sim\,\frac{l^2}{l_p^2}2\sqrt{6\pi\,N}(1-c^2).
	\label{action_X_b}
\end{equation}

We can now use the relation between $N$ and the proper length of the interval 

\begin{equation}
	N=\rho V=\rho\, \zeta_0^{(d)}\tau^d\Big\rvert_{d=4}=\rho\,\frac{\pi}{24}\tau^4
\end{equation}

and write the continuum limit of the action as 

\begin{equation}
	\lim_{\rho\rightarrow\infty}\frac{1}{\hbar}\left\langle S^{(4)}_{BD}(X)\right\rangle=\,\frac{1}{l_p^2}\,\pi\tau^2(1-c^2).
	\label{action_X_c}
\end{equation}

From this result we can easily infer the continuum limit of the action evaluated on the region $Y$. The BD action is invariant under time reversion, then when evaluated on the region $Y$, its value will only depend on the position of the horizon, i.e. $c\rightarrow1-c$. We infer

\begin{equation}
	\lim_{\rho\rightarrow\infty}\frac{1}{\hbar}\left\langle S^{(4)}_{BD}(Y)\right\rangle=\,\frac{1}{l_p^2}\,\pi\tau^2c(2-c).
	\label{action_Y}
\end{equation}

We are now able compute the SMI between the regions $X$ and $Y$. Recalling 

\begin{equation}
	\lim_{\rho\rightarrow\infty}\frac{1}{\hbar}\left\langle S^{(4)}_{BD}(\C)\right\rangle=\,\frac{1}{l_p^2}\,\pi\tau^2,
\end{equation}

one has, using \eqref{SMI_b}, 

\begin{equation}
	\left\langle \mathcal{I}^{(4)}[X,Y]\right\rangle=\frac{2}{l_p^2}\pi\tau^2c(1-c).
	\label{SMI_XY}
\end{equation}

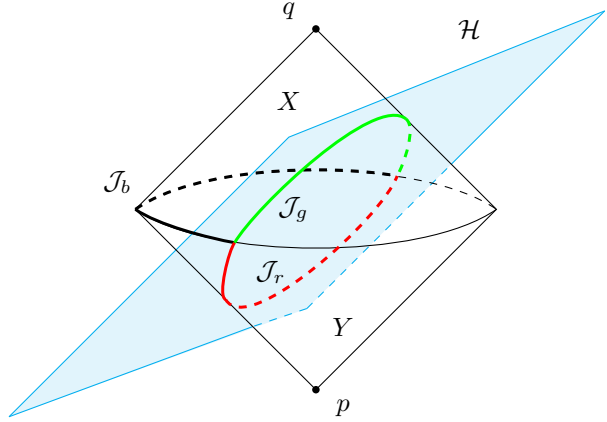
\begin{figure}[h]
	\centering 
	\begin{tikzpicture}[scale=1.2]
		
		\fill[cyan!10] (-0.1,-1.1) -- (3.2,2.2) -- (-0.3,.8) -- (-3.4,-2.3)--(-0.1,-1.1);
		\draw[cyan] (3.2,2.2) -- (-0.3,.8) -- (-3.4,-2.3);
		\draw[cyan] (3.2,2.2) -- (1.5,.5);
		\draw[cyan] (-.7,-1.3) -- (-3.4,-2.3);
		\draw[cyan, dashed] (-0.1,-1.1) -- (1.5,.5);
		\draw[cyan, dashed] (-0.1,-1.1) -- (-.7,-1.3);
		
		\draw[-] (0,-2) -- (2,0) -- (0,2) -- (-2,0) -- cycle;
		
		\node at (0,-2)[circle,fill,inner sep=1pt]{};
		\node at (0,2)[circle,fill,inner sep=1pt]{};
		\node at (-0.3,2.2) {$q$};
		\node at (0.3,-2.2) {$p$};
		
		\node at (-.3,1.2) {$X$};
		\node at (.3,-1.3) {$Y$};
		
		\node at (-0.5,-0.7) {$\J_r$};
		\node at (-0.25,0.0) {$\J_g$};
		\node at (-2.2,0.3) {$\J_b$};
		
		\node at (1.7,2) {$\mathcal{H}$};
		
		\draw (-2,0) to[out=-35,in=-145, distance=1cm ] (2,0);
		\draw[dashed] (-2,0) to[out=35,in=-215, distance=1cm ] (2,0);
		
		\draw[line width=0.4mm] (-2,0) to[out=-35,in=-190, distance=.3cm ] (-.9,-.37);
		\draw[line width=0.4mm, dashed] (-2,0) to[out=35,in=-190, distance=.8cm ] (.9,.37);
		
		\draw[line width=0.4mm, red] (-.9,-.37) to[out=230,in=-200, distance=.1cm] (-1,-1);
		\draw[line width=0.4mm, green] (1,1) to[out=140,in=60, distance=.45cm] (-.9,-.37);
		\draw[line width=0.4mm, dashed, green] (1,1) to[out=320,in=60, distance=.15cm] (.9,.37);
		\draw[line width=0.4mm, dashed,red] (-1,-1) to[out=300,in=250, distance=.5cm] (.9,.37);
		
	\end{tikzpicture}
	\caption{Three dimensional $I[p,q]$ diamond cut by an horizon $\h$. The black bold line is the null-null joint section contributing to the action of region $X$. The green bold arch is the cone-horizon joint section contributing in region $X$, whereas the red arch contributes in region $Y$. }
	\label{SMI_joints}
\end{figure}

One shall now compare this result to the volume of the dimension $2$ joint of the region $X$. This joint is composed of three sections: the intersection of the future light-cone of $p$ with the past light-cone of $q$ $\J_b$, and the intersections $\J_r,\J_g$ between both those light-cones and the horizon $\h$. These are named after their colour labels in Figure \ref{SMI_joints}.  \\

We start considering the cone-cone piece of the joint $\J_b$ (i.e the black bold curve in Figure \ref{SMI_joints}). For a diamond of proper length $\tau$, boosted such that it is spherically symmetric around the time direction, the intersection of the light-cones emanating from $p$ and $q$ is a sphere of radius $\tau/2$. Considering now the spacetime to be mapped by a chart $(x_0,x_1,r,\theta)$ centred in the middle of the diamond, one can parametrise the joint as the surface 

\begin{equation}
	\J_b(x_1,\theta)=\left(0,x_1,\sqrt{\left(\frac{\tau}{2}\right)^2-x_1^2}\sin\theta,\,\sqrt{\left(\frac{\tau}{2}\right)^2-x_1^2}\cos\theta\right).
	\label{cone-cone joint}
\end{equation}

One can then see the joint as a surface of revolution around the segment $x_1\in[-\tau/2,\tau/2]$. The truncation induced by the horizon limits the span of the coordinate $x_1$ to $x_1\in[-\tau/2,\tau/2-\sqrt{2}\lambda]$. Thus, the surface area of the joint is given by 

\begin{align}
	\mathrm{vol}\left(\J_b\right)&=\int_0^{2\pi}\drm\theta\int_{x_{1,min}}^{x_{1,max}}\drm x_1\,r(x_1)\sqrt{1+r'(x_1)^2},\nonumber\\
	&=\int_0^{2\pi}\drm\theta\int_{-\frac{\tau}{2}}^{\frac{\tau}{2}-\sqrt{2}\lambda}\drm x_1\,\sqrt{\left(\frac{\tau}{2}\right)^2-x_1^2}\sqrt{1+\left(\frac{x_1}{\sqrt{\left(\frac{\tau}{2}\right)^2-x_1^2}}\right)^2},\nonumber\\
	&=\pi\tau^2\left(1-\frac{\sqrt{2}\lambda}{\tau}\right)=\pi\tau^2\left(1-c\right).
	\label{cone-cone joint area}
\end{align}

The intersection of the light-cones with the horizon are dimension $2$ null surfaces in $\mathbb{M}^{4}$, in red and green on Figure \ref{SMI_joints}. In order to compute their surface area, we consider the coordinate chart $(t,x,y,z)$ on $\mathbb{M}$ and first look at the future light-cone $\mathcal{L}_p$, up to a time coordinate $t\leq\tau/2$. In this coordinate, the horizon is at $\h=\{t-x=\sqrt{2}\lambda\}$. We can therefore find a condition for $x$, i.e. $x=t-\sqrt{2}\lambda$. We can now impose the light-cone definition to the $4$-tuple $(t,t-\sqrt{2}\lambda,y,z)$

\begin{align}
	0&=-t^2+x^2+y^2+z^2 & &\Rightarrow & y^2+z^2=2\sqrt{2}t\lambda-2\lambda.
\end{align}  

We choose to parametrise the coordinates $(y,z)$ as a function of $(r,\theta)$ as $(y,z)\rightarrow(r\,\mathrm{sin}(\theta),r\,\mathrm{cos}(\theta))$. We then have $r^2=2\sqrt{2}t\lambda-2\lambda$ and the paraboloid (i.e the red curve in Figure \ref{SMI_joints}) is parametrized by

\begin{equation}
	\J_r:=\h\cap\mathcal{L}_p=\left(t,t-\sqrt{2}\lambda,\sqrt{2\sqrt{2}t\lambda-2\lambda},\mathrm{sin}(\theta),\sqrt{2\sqrt{2}t\lambda-2\lambda}\,\mathrm{cos}(\theta)\right).
\end{equation}

We shall now pull back the metric onto this hypersurface. Following \cite{poisson_2004} (3.2), we look for the hypersurface's tangent vectors

\begin{align}
	e^\alpha_t&=\frac{\partial x^\alpha}{\partial t}=\left(1,1,\frac{\sqrt{2}\lambda}{\sqrt{2\sqrt{2}t\lambda-2\lambda}}\,\mathrm{sin}(\theta),\frac{\sqrt{2}\lambda}{\sqrt{2\sqrt{2}t\lambda-2\lambda}}\,\mathrm{cos}(\theta)\right),\nonumber\\
	e^\alpha_\theta&=\frac{\partial x^\alpha}{\partial \theta}=\left(0,0,\sqrt{2\sqrt{2}t\lambda-2\lambda}\,\mathrm{cos}(\theta),-\sqrt{2\sqrt{2}t\lambda-2\lambda}\,\mathrm{sin}(\theta)\right).\nonumber\\
\end{align} 

Inner products between these vectors define the induced metric on $\h\cap\mathcal{L}_O$, $\sigma_{AB}$, with $\theta^A=(t,\theta)$. Thus, 

\begin{equation}
	\sigma_{AB}\drm\theta^A\drm\theta^B=\frac{2\lambda^2}{2\sqrt{2}t\lambda-2\lambda}\drm t^2+(2\sqrt{2}t\lambda-2\lambda)\drm\theta^2.
\end{equation}

The surface element is given by 

\begin{equation}
	\drm\Sigma=\sqrt{\sigma}\drm t\drm \theta=\sqrt{2}\lambda \drm t\drm\theta,
\end{equation}

And the red area becomes
\begin{equation}
    	\mathrm{vol}\left(\J_r\right)=\int_0^{2\pi}\drm\theta\int_{\frac{\sqrt{2}\lambda}{2}}^{\frac{\tau}{2}}\drm t\,\sqrt{2}\lambda=\pi \tau^2\,c(1-c),
\end{equation}

Now we need to consider also the intersection between the upper half of the diamond and the horizon $\J_g$, i.e. the green curve in Figure \ref{SMI_joints}. In this case, it would be a past-directed null cone with tip at $t=\tau$ and $t\in[\tau/2,\tau]$.  Note that the area of the intersection with one cone is invariant under $c\rightarrow 1-c$, so $\mathrm{vol}(\J_g)=\mathrm{vol}(\J_r)$. The joint area is therefore given by 
\begin{equation}
	\mathrm{vol}\left(\J\right)=2\mathrm{vol}\left(\J_r\right)=2\pi \tau^2\,c(1-c).
\end{equation}

Comparing this result to equation \eqref{SMI_XY}, we can see that the mutual information recovers precisely the area of the intersection between the the causal horizon and the null hypersurfaces defining the causal diamond. Thus, we obtain the claimed result \eqref{result2}. Although we present the calculations only for four dimensions, it should be straightforward to obtain the same conclusion in any dimensions.\\

This result, other than being an encouraging step in the understanding of SMI, acts as an independent check of the conjecture

\begin{equation}
	\lim_{\rho\rightarrow\infty}\frac{1}{\hbar}\Big\langle S_{BD}^{(d)}(I[p,q])\Big\rangle=\frac{1}{l_p^{d-2}}\int_{I[p,q]}\drm^dx\,\sqrt{-g}\frac{R}{2}+\frac{1}{l_p^{d-2}}\mathrm{vol}\left(\mathcal{J}^{(d-2)}\right)
	\label{Dowker_conjecture}
\end{equation}

about Benincasa-Dowker action's boundary terms \cite{buck2015boundary}. It states the continuum limit of the BD action, evaluated on a causal set generated from sprinkling on a compact region with null boundaries, is proportional to the Einstein-Hilbert action plus a joint term proportional to its co-dimension $2$ volume. This conjecture is recently verified for causal diamonds in perturbative regimes \cite{machet2020continuum,dowker2020boundary}. Here we consider the region $X$ with more complex joints than the simple light-cone-light-cone intersection. As depicted in Figure 8, besides the usual causal diamond joint (black), it consists of two extra pieces due to the intersection of the Rindler horizon (green and red). Our results here show that the BD action evaluated on the region $X$ (or $Y$ with an order-reversed argument) admits codimension-two boundary terms proportional to the area. \\

 More precisely, if we compared the volumes of $\J_r,\J_g$ with the action evalualted on $X$ \eqref{action_X_c} and $Y$ \eqref{action_Y}, we see only $\J_g$ and $\J_b$ contribute to the continuum limit of $S_{BD}(X)$, whereas only $J_r$ and $\J_b$ contribute to the continuum limit of $S_{BD}(Y)$. It actually follows nicely from the arguments in favour of $\eqref{Dowker_conjecture}$. The points sampled close to the part of $\J_r$ in the past always have enough spacetime in their future to contribute to the bulk term in the continuum limit. Thus one should not expect a boundary contribution from this joint when computing the BD action on $X$, and vice versa for $Y$. (See the introduction of \cite{dowker2020boundary} for details.)

\section{Conclusion}

The horizon molecules program is a viable way to define a kinematical entropy in CST with respect to a spacelike hypersurface crossing a causal horizon. We apply the recent proposal by Barton et al to null hypersurfaces. In a generic curved spacetime, the horizon molecule count can receive geometric contributions along the null segment away from the horizon-hypersurface intersection $\J$. Hence, the behaviour turns out deviating from the area law in the continuum limit whenever we have a null segment in the hypersurface crossing the horizon. As we alluded to in the introduction of section \ref{curved},  we also expect a deviation even in Minkowski spacetime when the null hypersurface possesses non-trivial extrinsic curvature in the vicinity of $\J$. We shall leave the verification of these cases to future works. We therefore need a patch to fix the horizon molecule proposal if possible. One can naively restore the area-law claimed in \cite{Dowker}, by somehow deforming away the null segment in the hypersurface simaltaneously while taking the continuum limit. For instance, we can set

\begin{equation}
\lambda, \Lambda \xrightarrow{l\rightarrow0}0.
\end{equation}  

or gradually deform the null segment to be spacelike to recover the claim

\begin{equation}
\lim_{l \rightarrow0}I_n^{(d)}(y;l,\Lambda, \lambda)=a_n^{(d)}.
\label{claim}
\end{equation}

It is of course unphysical to deform the hypersurface by hand, so it must be realized inherently on causal sets in a natural way. However, we find it challenging to come up with a modified definition of horizon molecules such that it naturally incorporates this feature.\\

It would be useful to actually compute the horizon molecule count in the Schwarzschild spacetime, where we know there should be a unique sensible answer given by the area of the sphere at the Schwarzschild radius \footnote{We thank Ian Jubb for pointing this out.}. Therefore, physically we expect the result to be independent of the hypersurface chosen, whether it is null or spacelike. Confirming this behaviour will be an important sanity check for the horizon molecule proposal in general.\\

We then looked to the SMI definition and we computed it on some causal diamonds truncated by a causal horizon in Minkowski spacetime. We showed the SMI localizes to the intersection between the horizon and the boundaries of the causal diamond and scales proportionally to the intersections' area in the continuum limit. This represents first steps towards an analytic understanding of this quantity and its relation to horizon entropy. Moreover, we obtained a non-trivial check of Benincasa-Dowker action behaviour on a causally convex region with null hypersurfaces joints different from the ones of the standard causal diamond case. It is crucial to test the SMI proposal on null hypersurfaces which are not causal horizons, in order to check if the effective localisation of this quantity is proper to globally defined horizons or if it is also present in other constructions. The final hope is for one to find a causal set entropy that could lead to a Generalised Second Law of black hole thermodynamics. This connects more widely to the incorporation of matter and fields in CST and to their dynamics in a horizon spacetime. This therefore motivates us to look for a connection between SMI/horizon molecules and the study of entanglement entropy in CST \cite{Sorkin_2018,chen2020towards,surya2020entanglement}.

\ack

We thank Ian Jubb for valuable feedback and discussions, and Fay Dowker for introducing the problem of skinny causal diamond volume to us. This work was supported by the ESA Prodex project 'LISA EMRI/IMRAC waveform modelling' PEA 4000131558, the C16/16/005 research grant of the KU Leuven and the FWO Grant No. G092617N.

%\appendix
%\section{A}

\section*{References}
\bibliographystyle{unsrt}

\bibliography{References}

\begin{thebibliography}{10}

\bibitem{Hawking}
S~W Hawking.
\newblock
  \href{https://link.aps.org/doi/10.1103/PhysRevLett.26.1344}{Gravitational
  Radiation from Colliding Black Holes}.
\newblock {\em Phys. Rev. Lett.}, 26:1344--1346, May 1971.

\bibitem{Bekenstein}
J~D Bekenstein.
\newblock \href{https://link.aps.org/doi/10.1103/PhysRevD.7.2333}{Black Holes
  and Entropy}.
\newblock {\em Phys. Rev. D}, 7:2333--2346, Apr 1973.

\bibitem{JP}
T~Jacobson and R~Parentani.
\newblock \href{https://doi.org/10.1023/A:1023785123428}{Horizon Entropy}.
\newblock {\em Foundations of Physics}, 33(2):323--348, 2003.

\bibitem{Surya}
S~Surya.
\newblock \href{https://doi.org/10.1007/s41114-019-0023-1}{The causal set
  approach to quantum gravity}.
\newblock {\em Living Reviews in Relativity}, 22(1), Sep 2019.

\bibitem{DS}
D~Dou and R~D Sorkin.
\newblock \href{https://doi.org/10.1023/A:1023781022519}{Black-Hole Entropy as
  Causal Links}.
\newblock {\em Foundations of Physics}, 33(2):279--296, 2003.

\bibitem{Marr}
S~K Marr.
\newblock \href
  {https://spiral.imperial.ac.uk/bitstream/10044/1/11818/2/Marr-SK-2007-PhD-Thesis.pdf}{Black
  hole entropy from causal sets}.
\newblock {\em Imperial College London, PhD thesis}, August 2007.

\bibitem{Dowker}
C~Barton, A~Counsell, F~Dowker, D~S~W Gould, G~Taylor, and I~Jubb.
\newblock \href{{https://link.aps.org/doi/10.1103/PhysRevD.100.126008}}{Horizon
  molecules in causal set theory}.
\newblock {\em Phys. Rev. D}, 100:126008, Dec 2019.

\bibitem{Sorkin_2018}
R~D Sorkin and Y~K Yazdi.
\newblock \href{http://dx.doi.org/10.1088/1361-6382/aab06f}{Entanglement
  entropy in causal set theory}.
\newblock {\em Classical and Quantum Gravity}, 35(7):074004, Mar 2018.

\bibitem{chen2020towards}
Y~Chen, L~Hackl, R~Kunjwal, H~Moradi, Y~K. Yazdi, and Miguel Zilh{\~a}o.
\newblock \href{https://doi.org/10.1007/JHEP11(2020)114}{Towards spacetime
  entanglement entropy for interacting theories}.
\newblock {\em Journal of High Energy Physics}, 2020(11):114, 2020.

\bibitem{surya2020entanglement}
S~Surya, Y~K Yazdi, and et. al.
\newblock \href{https://arxiv.org/abs/2008.07697}{Entanglement Entropy of
  Causal Set de Sitter Horizons}.
\newblock {\em arXiv preprint arXiv:2008.07697}, 2020.

\bibitem{buck2015boundary}
M~Buck, F~Dowker, I~Jubb, and S~Surya.
\newblock
  \href{https://iopscience.iop.org/article/10.1088/0264-9381/32/20/205004/meta}{Boundary
  terms for causal sets}.
\newblock {\em Classical and Quantum Gravity}, 32(20):205004, 2015.

\bibitem{Cauchy}
P~T Chrusciel and T-T Paetz.
\newblock \href{https://arxiv.org/abs/1203.4534}{The Many ways of the
  characteristic Cauchy problem}.
\newblock {\em Class. Quant. Grav.}, 29:145006, 2012.

\bibitem{GNC}
H~Friedrich, I~I Racz, and R~M Wald.
\newblock \href{https://link.springer.com/article/10.1007\%2Fs002200050662}{On
  the Rigidity Theorem for Spacetimes with a Stationary Event Horizon or a
  Compact Cauchy Horizon}.
\newblock {\em Communications in Mathematical Physics}, 204:691--707, 1999.

\bibitem{Benincasa}
D~Benincasa.
\newblock
  \href{https://spiral.imperial.ac.uk/bitstream/10044/1/14170/1/Benincasa-DMT-2013-PhD-Thesis.pdf}{The
  Action of a Causal Set}.
\newblock {\em Imperial College London, PhD thesis}, March 2013.

\bibitem{benincasa2010scalar}
D~Benincasa and F~Dowker.
\newblock
  \href{https://journals.aps.org/prl/abstract/10.1103/PhysRevLett.104.181301}{Scalar
  curvature of a causal set}.
\newblock {\em Physical review letters}, 104(18):181301, 2010.

\bibitem{dowker2013causal}
F~Dowker and L~Glaser.
\newblock
  \href{https://iopscience.iop.org/article/10.1088/0264-9381/30/19/195016}{Causal
  set d'Alembertians for various dimensions}.
\newblock {\em Classical and Quantum Gravity}, 30(19):195016, 2013.

\bibitem{poisson_2004}
E~Poisson.
\newblock {\em
  \href{https://www.cambridge.org/core/books/relativists-toolkit/DA7ED68B971708A0F782257F948981E7}{A
  Relativist's Toolkit: The Mathematics of Black-Hole Mechanics}}.
\newblock Cambridge University Press, 2004.

\bibitem{machet2020continuum}
L~Machet and J~Wang.
\newblock \href{https://iopscience.iop.org/article/10.1088/1361-6382/abc274}{On
  the continuum limit of Benincasa-Dowker-Glaser causal set action}.
\newblock {\em Classical and Quantum Gravity}, 38:015010, 2021.

\bibitem{dowker2020boundary}
F~Dowker.
\newblock
  \href{https://iopscience.iop.org/article/10.1088/1361-6382/abc2fd/meta}{Boundary
  contributions in the causal set action}.
\newblock {\em Classical and Quantum Gravity}, 2020.

\end{thebibliography}

\appendix
\section{Volume of a skinny causal diamond}

\subsection{Introduction}

Given a causal diamond $I(p,q)$, one can shrink it down and express its volume as a perturbative expansion, which depends on how the interval is shrunk and the data associated with the limit. We consider here a particular small volume limit of the interval where the geodesic between $p,q$ tends toward an arbitrary null curve $\gamma$ on the lower light-cone, as $q$ approaches the corner of the causal diamond. When the volume gets small, the causal diamond becomes a long skinny interval and its volume expansion is determined by the geometric data on $\gamma$. By setting up the integral in null Fermi Normal Coordinates (NFNC) along $\gamma$, we give here a systematical calculation of the volume expansion for any dimensions.\\

More precisely, given a null curve $\gamma(p,o)$ of affine parameter distance $L$, we shoot a short null geodesic $\gamma'(o,q)$ that is not parallel to $\dot{\gamma}$ at $o$. $\gamma(o,q)'$ is parametrized by $\lambda\in [0,s]$ where $s$ is assumed to be very small such that the proper time $\tau(p,q)$ is small. For large curvatures or long stretched-out interals, one might have caustics on the light-cone so that the geometry of the interval $I(p,q)$ becomes complicated. Therefore, we assume that the curvature scale $1/\sqrt{R}$ is much larger than $L$, i.e. $RL^2\ll 1.$ The task is to evaluate the spacetime volume of $I(p,q)$ perturbatively using the curvature data on $\gamma(p,o)$.\\

The resulting volume expansion in $s$ have coefficients in terms of the integrals of curvature data along $\gamma(p,o)$. In particular, the result we obtain for spacetime dimension $n=4$ is
\begin{align}\label{volumeformula}
V^{(4)}=& \tau^4\pi\left(\frac{1}{24}+\frac12\int_0^L\dd X^+ \frac{(L-X^+)^2}{L^3X^+}\int_0^{X^+} \dd x \int_0^{x} \dd x'\;x'  R\indices{_+_+}(x')+O(L^3)\right)+O(s^3).
\end{align}
where $\tau=\sqrt{sL}+O(s^2L^2)$ is the propertime between $p,q$. The volume is expanded both in $s$ and $L$.\\

Throughout, we use Greek indices $\al,\be,\mu,\dots$ for spacetime NFNC coordiantes, while Latin indices $\a,\b,\c,\dots$ are codimension-1 indices on the subspace orthogonal to $\gamma$ and the Lain indices $i,j,k,\dots$ are codimension-2 indices for the transverse spatial dimensions. Index $+$ denotes the direction of $\gamma(p,o)$. We will leave out the Big $O$ notations for the expansions and expand up to the leading perturbative order. 
\subsection{The volume in the Minkowski spacetime }

Let us first look at the problem in the Minkowski spacetime. The answer is known to be
\beq
V_0 = \frac{\Omega_{n-2}\tau^n}{2^{n-1}n(n-1)},    \;\;\;\;\;\;\;\;\text{where}\;\; \Omega_{n-2}=\frac{2\pi^{\frac{n-1}{2}}}{\Gamma(\frac{n-1}{2})}.
\eeq
The easiest way to compute it is to boost the tilted long skinny interval to an upright configuration and integrate in that frame. However, it is tricky to adopt the same strategy in a curved spacetime without global Lorentzian symmetry. Hence, it is worth a detour to set up the NFNC along $\gamma(p,o)$ and perform the integration in the tilted interval. It will give some guidance on how to set up the integral in the general case.

We shall denote the null generator of $\gamma(p,o)$ ($\gamma$ for short) as $\ell^+$ and the ingoing null vector which generates $\gamma(o,q)$ as $\ell^-$. We choose the normalisation $\ell^-\cdot \ell^+=-2$ at $o$ instead of $-1$ for convenience in calculations later. $\ell^-$ is parallelly transported along $\gamma$ so both $\ell^\pm$ is defined on $\gamma$ and we do not need their extensions outside $\gamma$. In NFNC, we choose 
\beq
\ell^{+\mu}=(1,0,0) = (1,0,0,n^i),\;\;\;\ell^{-\mu}=(0,1,0)=(0,1,0,n^i)
\eeq
where the first coordinate is written in the form of $(X^+, X^-, X^i)$ and the second has the spatial transverse part represented in spherical coordinates $(X^+,X^-,r,\theta^i)$.

The metric reads
\beq
\eta=-2\dd X^+\dd X^-+\delta_{ij}\dd X^i\dd X^j = -2\dd X^+\dd X^-+\dd r^2+r^{n-3}\dd \Omega_{n-3}.
\eeq
$p$ is the origin, $o$ has coordinates $(L,0,0)$, and $q$ is chosen to be located at $X^\mu(p)=(L,s,0)$. The boundary of the causal diamond is given by the following constraint equations
\beq
X^-_\vee = \frac{r^2}{2X^+},\;\;X^-_\wedge = \frac{r^2}{2(X^+-L)}+s\label{flatcons}
\eeq
where $r^2:=X^iX_i$ and we see that each conic section of fixed $X^+$ is indeed a parabola. The proper time between $p,q$ is given by 
\beq
\tau=\sqrt{2Ls}
\eeq

We can thus set up the integral in the following way
\begin{align}
V=&\int_0^L \dd X^+\int_{S^{n-3}}\dd\Omega_{n-3}\int_0^{r^*}r^{n-3}\dd r\int_{X_\vee^-}^{X_\wedge^-}\dd X^-\sqrt{\eta}
\end{align}
where $r^*:=\sqrt{2sX^+(1-X^+/L)}$ locates the intersection between $X^-_\vee$ and $X^-_\wedge$.

Hence, we have
\beq
\begin{aligned}
V=&\int_0^L \dd X^+\int_{S^{n-3}}\dd\Omega_{n-3}\int_0^{\sqrt{2sX^+(1-X^+/L)}}  \dd r\;r^{n-3} \left(s+\frac{Lr^2}{2X^+(X^+-L)}\right),\\
=&\int_0^L \dd X^+\int_{S^{n-3}}\dd\Omega_{n-3}\; \frac{(2s)^\frac{n}{2}[X^+(1-X^+/L)]^{\frac{n}{2}-1}}{n(n-2)},\\
=&\frac{ \Omega_{n-3}(2s)^\frac{n}{2}}{n(n-2)}\int_0^L \dd X^+\; [X^+(1-X^+/L)]^{\frac{n}{2}-1},\\
=& \frac{ \Omega_{n-3}(2sL)^\frac{n}{2}}{n(n-2)}\frac{\Gamma(\frac{n}{2})^2}{\Gamma(n)} = \frac{\Omega_{n-2}2^{1-n}\tau^n}{n(n-1)} = V_0.
\end{aligned}
\eeq
This is consistent with the known result.

\subsection{The boundary of the causal diamond}
\emph{Perturbative order}: We are interested in the volume expansion in small parameter $s$ and also assume the curvature is much smaller than $L$, the transverse span of the causal diamond is also small. We shall keep only first order curvature terms in our expansions, that is, we expand any dimensionless quantities up to order $O(X^2)$ where $X$ represents any coordinates, and dimensionful quantities such as $X^+$ to order $O(X^3)$, etc.

In a curved spacetime, we can decompose the metric as 
\beq
\begin{aligned}
g=&\eta+h\,,\\
=&-2\dd X^+\dd X^-+\delta_{ij}\dd X^i\dd X^j \;\;\;(=\eta) \\
&-R_{+\a +\b}\dd X^{+2}X^{\a}X^{\b}-\frac43 R_{+\a\b\c}X^{\a}X^{\c}\dd X^{\b}\dd X^+-\frac13 R_{\a\b\c\d}X^{\b}X^{\d} \dd X^{\a}\dd X^{\c}\;\;\;(:=h)\\
&+O(X^3).
\end{aligned}
\eeq

The volume form in regions close to $\gamma$ in NFNC is given by
\begin{align}
\sqrt{g}=&\sqrt{\eta}+\frac12 \eta^{\al\be}h_{\al\be}=1+\left(\frac13 R_{+\a -\b}-\frac16 R_{\a\b}\right)X^{\a} X^{\b},\nonumber \\
=&1 + \frac13 R_{+i-j}X^iX^j-\frac16 \left(R_{--}X^-X^-+R_{ij}X^iX^j\right)+O(X^3)  \label{vform}
\end{align}
where we've omitted the contribution with one $X^i$ as eventually these terms will vanish after the integral $\int \dd\Omega$. We shall also omit all such scalar terms with odd number of $X^i$'s in the following calculations for the same reason.

The Christoffel symbols vanish on $\gamma$ but their derivatives do not. The non-zero components evaluated on $\gamma$ are
\beq
\Gamma\indices{^\mu_{\al +,\be}}=R\indices{^\mu_{\al\be +}}\;\;\;, \;\;\;\;\Gamma\indices{^\mu_{\a\b ,\c}} = -\frac23R\indices{^\mu_{(\a\b)\c}}\label{chris}.
\eeq
Near $\gamma$ we can expand Christoffel symbols as
\beq
\Gamma\indices{^\mu_{\al\be}}(X)= \Gamma\indices{^\mu_{\al\be}}\Big |_\gamma + \Gamma\indices{^\mu_{\al\be, \nu}}\Big |_\gamma X^\nu = \Gamma\indices{^\mu_{\al\be, \c}}\Big |_\gamma X^{\c}+O(X^2)
\eeq
where the second equality is due to $\Gamma\indices{^\mu_{\al\be, +}}=0$. 
Although the Christoffel symbols are not necessarily small, the infinitesimal comes from the $X^c$ as the ingoing null direction $X^-$ and the transverse direction $X^i$ are of length order $s$ and $\sqrt{s}$ respectively, as inferred from the flat spacetime calcula. This justifies the perturbative expansion.

The interval is placed at the same coordinates as in the flat case, namely, 
\beq
X^\mu(p)=(0,0,0), X^\mu(q)=(L,s,0). 
\eeq
For convenience, we also define
\beq
Y^\mu=(L,s,0)\,.
\eeq
which is the coordinate of $q$ in the Minkowski spacetime.

In a curved spacetime, the boundary of the long skinny interval is distorted by the curvature. To quantitatively compute the perturbations of the light-cone, we shall impose the geodesic eqaution with the connection data expanded around $\gamma(\lambda)$, which is parameterized by some affine parameter $\lambda$. 

Null geodesics imposing the following constraints
\beq
g(\la)_{\al\be}\dot{X}^\al(\lambda) \dot{X}^\be(\lambda) = 0,\;\;\;\; \ddot{X}^\mu(\lambda) + \Gamma\indices{^\mu_\al_\be}\dot{X}^\al(\lambda) \dot{X}^\be(\lambda) = 0. \label{constraints}
\eeq
The tangents on the lower light-cone, denoted as $\dot{X}_\vee$, can be determined to the leading perturbative order 
\beq
\dot{X}_\vee^\mu(\lambda) = X_0^\mu - \int_0^{\la} \dd\ta\; \ta\;\Gamma\indices{^\mu_{\al\be,c}}(\ta)X_0^c X_0^\al X_0^\be+O(X^4)
\eeq
where $X_0^\mu$, a null vector w.r.t. $\eta$, is the initial direction of a geodesic.

Similarly, on the upper light-cone, one has
\beq
\dot{X}_\wedge^\mu(\lambda) = - X_0^\mu +\frac12 \Gamma\indices{^{\mu}_{\al\b,\c}}X_0^\al Y^{\b}Y^{\c} - \int_0^{\la} \dd\ta\;\Gamma\indices{^\mu_{\al\be,c}}(\ta)(Y^{\c}-\ta X_0^{\c}) X_0^\al X_0^\be +O(X^4).
\eeq

Then coordinates of along a null geodesic on the lower/upper light-cone are given by
\begin{align}
X_\vee^\mu(\la) =& \la X_0^\mu- \int_0^{\la}\dd \la ' \int_0^{\la '} \dd\ta\; \ta\;\Gamma\indices{^\mu_{\al\be,c}}X_0^c X_0^\al X_0^\be +O(X^4),\label{lowercoords}\\
X_\wedge^\mu(\la) =& Y^\mu-\la X_0^\mu +\frac12 \int_0^\la \dd\la '\Gamma\indices{^{\mu}_{\al\b,\c}}X_0^\al Y^{\b}Y^{\c}-\int_0^{\la}\dd \la ' \int_0^{\la} \dd\ta\;\Gamma\indices{^\mu_{\al\be,c}}(Y^{\c}-\ta X_0^{\c}) X_0^\al X_0^\be +O(X^4)\label{uppercoords}
\end{align}
where the boundary conditions are set to be $X_\vee^\mu(0)=0,X_\wedge^\mu(0)=Y^\mu.$ We henceforth leave out the arguments of $\Gamma$'s for convenience and its dependence on $\la$ is implicitly understood.  We can see that the leading terms are the coordinates in the Minkowski spacetime. 

As in the flat case (\ref{flatcons}), we would like to have the deparameterized versions of $X_{\vee,\wedge}^-(X^-,X^i)$. According to (\ref{chris}), the terms involving the affine connection expand as
\beq
\begin{aligned}
\Gamma\indices{^+_{\al\be,\c}}X_0^{\c}X_0^\al X_0^\be =&\,-R_{-+-+}X_0^-X_0^{+2}+2R_{-i+j}X_0^i X_0^j X_0^+ +O(X^4)\,,\\
\Gamma\indices{^-_{\al\be,\c}}X_0^{\c}X_0^\al X_0^\be =&\,%\;2R\indices{^-_\al_\c_+}X_0^{\c} X_0^\al X_0^+=
2R\indices{_+_-_+_-}X_0^{-2} X_0^+ +2R\indices{_+_i_+_j}X_0^iX_0^j X_0^+ +O(X^4)\,,\\
\Gamma\indices{^i_{\al\be,\c}}X_0^{\c}X_0^\al X_0^\be X_{0\,i} =&\;R_{i+j+}X_0^iX_0^jX_0^{+2} +2R_{i-j+}X_0^i X_0^jX_0^-X_0^+ +O(X^5)\,,\\
\Gamma\indices{^+_{\al\b,\c}}X_0^\al Y^{\b}Y^{\c}=&\,%s^2R\indices{^+_{--+}}X_0^+ 
0 +O(X^4)\,,\\
\Gamma\indices{^-_{\al\b,\c}}X_0^\al Y^{\b}Y^{\c}=&\,%s^2R\indices{^-_{--+}}X_0^+=
s^2R\indices{_{+-+-}}X_0^+ +O(X^4)\,,\\
\Gamma\indices{^{\mu}_{\al\b,\c}}X_0^\al Y^{\b}Y^{\c}X_{0\,\mu}=&\,%s^2R\indices{^{\mu}_{--+}}X_0^+X_{0\mu}-\frac13 s^2R\indices{^{\mu}_{-j-}}X_0^jX_{0\mu}=
-s^2R\indices{_{+-+-}}X_0^{+2}-\frac13 s^2 R\indices{_{i-j-}}X_0^iX_0^j +O(X^5)\,.
\end{aligned}
\eeq

We shall now demonstrate the calculation explicitly for the lower cone. After substitution to (\ref{lowercoords},\ref{uppercoords}), we obtain
\begin{align}
X_\vee^+(\la) =& \la X_0^+ + \int_0^{\la}\dd \la '\int_0^{\la '} \dd \ta \;\ta \left( R_{-+-+}X_0^-X_0^{+}-2R_{-i+j}X_0^i X_0^j \right)X_0^+ +O(X^4),\label{Xplus}\\
X_\vee^-(\la) =& \la X_0^- -  \int_0^{\la}\dd \la '\int_0^{\la '} \dd \ta \;\ta \left(2R\indices{_+_-_+_-}X_0^{-2} +2R\indices{_+_i_+_j}X_0^iX_0^j\right)X_0^+ +O(X^4), \label{Xminus}\\
r^2(\la):=& X_{\vee\,i}X_\vee^i = \la^2 r_0^2 - \la \int_0^{\la}\dd \la '\int_0^{\la '} \dd \ta \;\ta \left(2R_{i+j+}X_0^+ + 4R_{i-j+}X_0^-\right)X_0^iX_0^jX_0^+  +O(X^5).\label{Xi}
\end{align}

We know that $X_0^-=\frac{r_0^2}{2X_0^+}$, and together with (\ref{Xplus},\ref{Xi}) we have the boundary described by
\begin{align}
\la X_0^- = \frac{\la r_0^2}{2X_0^+} =& \frac{\la}{2X_0^+}\left(\frac{r^2}{\la^2}+\frac{1}{\la} \int_0^{\la}\dd \la '\int_0^{\la '} \dd \ta \;\ta \left(2R_{i+j+}X_0^iX_0^jX_0^{+2} + 4R_{i-j+}X_0^i X_0^jX_0^-X_0^+\right)\right),\nonumber\\
=&\frac{r^2}{2X^+}+ \int_0^{\la}\dd \la '\int_0^{\la '} \dd \ta \;\ta \left(R_{i+j+}X_0^iX_0^jX_0^{+} + 2R_{i-j+}X_0^i X_0^jX_0^-\right),\nonumber\\
=&\frac{r^2}{2 X^+}+\frac{r^2}{2X^{+2}} \int_0^{\la}\dd \la '\int_0^{\la '} \dd \ta \;\ta \left( R_{-+-+}X_0^-X_0^{+2}-2R_{-i+j}X_0^i X_0^jX_0^+ \right) \nonumber \\
&+ \int_0^{\la}\dd \la '\int_0^{\la '} \dd \ta \;\ta \left(R_{i+j+}X_0^iX_0^jX_0^{+} + 2R_{i-j+}X_0^i X_0^jX_0^-\right)+O(X^4)
\end{align}
where we have used  (\ref{Xi}), (\ref{Xplus}) in the second and third equality respectively.

Altogether with (\ref{Xminus}), we have
\begin{align}
X_\vee^-(\lambda)=&\frac{r^2}{2 X^+}+\frac{r^2}{ 2X^{+2}} \int_0^{\la}\dd \la '\int_0^{\la '} \dd \ta \;\ta \left( R_{-+-+}X_0^-X_0^{+2}-2R_{-i+j}X_0^i X_0^jX_0^+ \right) \nonumber \\
&+ \int_0^{\la}\dd \la '\int_0^{\la '} \dd \ta \;\ta \left(R_{i+j+}X_0^iX_0^jX_0^{+} + 2R_{i-j+}X_0^i X_0^jX_0^-\right) \nonumber\\
&-2 \int_0^{\la}\dd \la '\int_0^{\la '} \dd \ta  \;\ta \left(R\indices{_+_-_+_-}X_0^{+}X_0^{-2} +R\indices{_+_i_+_j}X_0^iX_0^jX_0^+\right)+O(X^4)\,.
\end{align}

This is not good enough, we need to write $X^-$ in terms of $X^+, X^i$ and change the parameter $\lambda$ to $X^+$ so to obtain a constraint between the coordinates.
\beq\label{eq:constraintdown}
\begin{aligned}
X_\vee^- =&\;\frac{r^2}{2 X^+}-\frac{X^iX^j}{X^{+\,2}}\int_0^{X^+} \dd x \int_0^{x} \dd x'\;x'  R\indices{_i_+_j_+}-\frac{r^4}{4X^{+\,4}}\int_0^{X^+} \dd x \int_0^{x} \dd x'\;x' R_{-+-+}+O(X^4).
\end{aligned}
\eeq
where in the second equality we have repalce every $X^-$ on the RHS by $\frac{r^2}{2X^{+}}$, which holds at zeroth order. We see that the lower cone boundary is not perturbed at this order in NFNC. This is not the case for the upper cone boundary.

The same procedure gives the $X^-$ for the upper light-cone
\begin{align}\label{eq:constraintup}
X_\wedge^-=&s+\frac{r^2}{2(X^+-L)}-\frac{X^iX^j(X^+-L)}{X^{+\,3}}\int_0^{X^+} \dd x \int_0^{x} \dd x'\;x'  R\indices{_i_+_j_+}\nonumber\\
&-\frac{r^4}{4(X^+-L)X^{+\,3}}\int_0^{X^+} \dd x \int_0^{x} \dd x'\;x' R_{-+-+}+O(X^4).
\end{align}

%+\frac{s^2r^2n^in^j}{6(L-X^+)}\int_{X^+}^L \dd X^+\;R_{i-j-}(X^+) .

We leave out the calculation details here, which are more tedius than the lower cone case, but we can see that the final expression is simple and the perturbation terms are essentially due to the third term in (\ref{uppercoords}).

Equating $X_\vee^-=X_\wedge^-$ gives the constraint on the corner and thus the integral domain of $\int \dd r$. 
\begin{align}
r^{*2}=&\frac{2s(L-X^+)X^+}{L}+\frac{4s(L-X^+)^2}{X^+L}\int_0^{X^+} \dd x \int_0^{x} \dd x'\;x'  R\indices{_i_+_j_+}n^in^j\nonumber\\
&+\frac{2s^2(L-X^+)^2}{X^+L^2}\int_0^{X^+} \dd x \int_0^{x} \dd x'\;x' R_{-+-+}+O(X^5).
\end{align}

\subsection{The volume in a curved spacetime }
\subsubsection{The interval length}
First of all, we need to compute the proper time difference between $p,q$. The tangent on the geodesic $\gamma(p,q)$ reads
\begin{align}
\dot{X}^\mu(\la) =& Y^\mu+\int_0^1\dd \la '\int_0^{\la'}\dd \ta\; \ta \;\Gamma\indices{^\mu_{\al\be,\c}}Y^{\c}Y^\al Y^\be -\int_0^{\la}\dd \ta\; \ta \;\Gamma\indices{^\mu_{\al\be,\c}}Y^{\c}Y^\al Y^\be+O(X^4).
\end{align}
It clearly satisfies the geodesic equation.
The geodesic is 
\beq
X^\mu(\la)=\la Y^\mu+\la\int_0^1\dd \la '\int_0^{\la'}\dd \ta\; \ta \;\Gamma\indices{^\mu_{\al\be,\c}}Y^{\c}Y^\al Y^\be -\int_0^\la\dd \la '\int_0^{\la'}\dd \ta\; \ta \;\Gamma\indices{^\mu_{\al\be,\c}}Y^{\c}Y^\al Y^\be+O(X^4),
\eeq
which satisfies $X^\mu(0)=(0,0,0),\;X^\mu(1)=(L,s,0).$

The tangents are not normalized,
\begin{align}
\dot{X}^2=-2Ls+4(Ls)^2\int_0^\la \dd \ta\;\ta\;R_{+-+-}-4(Ls)^2\int_0^1\dd \la'\int_0^{\la'} \dd \ta\;\ta\;R_{+-+-}+O(X^5),
\end{align}
where $O(X^5)$ means terms with order $O(s^3L^2), O(s^2L^3)$ etc.

Since the affine parameter $\la$ runs from $0$ to $1$, the proper time is 
\beq
\tau=\int^1_0\dd \la  |\dot{X}|=\sqrt{2Ls}+O(X^4). \label{propertime}
\eeq
We see that at this perturbative order, the interval length is the same as in flat spacetime. 

\subsubsection{The volume integral}
Now we are ready to do the integration
\begin{align}
V=&\int_0^L \dd X^+\int_{S^{n-3}}\dd\Omega_{n-3}\int_0^{r^*}r^{n-3}\dd r\int_{X_\vee^-}^{X_\wedge^-}\dd X^-\sqrt{g}.\label{finalintegral}
\end{align}
In the calculation, we would like to collect terms in powers of $s$ and keep all terms up to $O(L^2)$. This is consistent with the perturbative order we are working with so far.

In order to tidy up the following calculations, we introduce some shorthands here.
\beq \label{shorthands}
\begin{aligned}
I_{ij}(X^+):=&\int_0^{X^+} \dd x \int_0^{x} \dd x'\;x'  R\indices{_i_+_j_+}\sim O(L^3), \;\;\;\;\;\;\;I(X^+):=I_{ij}n^in^j\sim O(L^3),\\
I_1(X^+):=&\int_0^{X^+} \dd x \int_0^{x} \dd x'\;x' R_{-+-+}\sim O(L^3)\;,\;\;\;\;\;\chi(X^+):= X^+(L-X^+)\sim O(L^2),\\
I_2(X^+):=&\int_0^{X^+} \dd x \int_0^{x} \dd x'\;x'  R\indices{_+_+}\sim O(L^3).
\end{aligned}
\eeq
where we also indicate its order in $L$.

In terms of these, we can rewrite the boundaries of the light-cone as
\beq
\begin{aligned}
X_\wedge^-=&s-\frac{r^2}{2(L-X^+)}+\frac{r^2(L-X^+)}{X^{+\,3}}I+\frac{r^4I_1}{4(L-X^+)X^{+\,3}}+O(X^4),\\
X_\vee^- =&\;\frac{r^2}{2 X^+}-\frac{r^2}{X^{+\,2}}I-\frac{r^4I_1}{4X^{+\,4}}+O(X^4),\\
X_\wedge^- - X_\vee^- =& s-\frac{r^2L}{2\chi}+\frac{r^2LI}{X^{+\,3}}+\frac{r^4LI_1}{4X^{+\,3}\chi}+O(X^4),\\
r^{*2}=&\frac{2s\chi}{L}+\frac{4s\chi^2I}{X^{+\,3}L}+\frac{2s^2\chi^2I_1}{X^{+\,3}L^2}+O(X^5).
\end{aligned}
\eeq
We shall do the nested integral (\ref{finalintegral}) one by one
\begin{align}
S_{X^-}:=&\int_{X_\vee^-}^{X_\wedge^-}\dd X^-\sqrt{g},\nonumber\\
=& \left(s-\frac{r^2L}{2\chi}+\frac{r^2LI}{X^{+\,3}}+\frac{r^4LI_1}{4X^{+\,3}\chi}\right)\left(1+ \frac13 R_{+i-j}X^iX^j-\frac16R_{ij}X^iX^j\right)+O(X^5).\label{x-integral}
\end{align}

Note we've left out the contribution $R_{--}X^{-2}$ in $\sqrt{g}$ (\ref{vform}), which is of order $O(s^3)$. 

Although in principle one perform the integral in arbitrary dimensions, it is complicated to write a closed-form formula for them. Instead, we focus on the case of $n=4$ from now on. Let's just focus on the leading term first. For four dimensions, that is with order $O(s^2L^2).$
The integral over the radial direction $r$ gives
\beq
\begin{aligned}
S^{(4)}_{r}:=&\int_0^{r^{*}}r S_{X^-}\dd r\;,\\
=&\int_0^{r^{*}}\dd r\,r^3\left(\frac{s}{r^2}-\frac{L}{2\chi}+\frac{LI}{X^{+\,3}}\right)+r^5\left(\frac{LI_1}{4X^{+\,3}\chi}+\frac{LI}{X^{+\,3}}+\left[\frac{s}{r^2}-\frac{L}{2\chi}\right]\left[\frac{1}{3}R_{+i-j}n^in^j-\frac{1}{6}R_{ij}n^in^j\right]\right),\\
=&\frac{sr^{*2}}{2}+\frac{r^{* 4}}{4}\left(\frac{LI}{X^{+\,3}}-\frac{L}{2\chi}\right)+O(s^3),\\
=&s^2\left(\frac{2\chi}{L}+\frac{4\chi^2I}{X^{+\,3}L}+\frac{2s\chi^2I_1}{X^{+\,3}L^2}\right)\left(\frac{1}{2}+\frac{\chi I}{X^{+\,3}}+\frac{s\chi I_1}{4X^{+\,3}L}\right)+O(s^3),\\
=&s^2\frac{\chi}{L}(1+\frac{2\chi I}{X^{+\,3}})^2+O(s^3).
\end{aligned}
\eeq

The spherical integral gives
\begin{align}
S^{(4)}_\Omega:=&\int_{S^1}\dd\Omega_1 S_r^{(4)}= s^2\pi\frac{\chi}{L}\left(1+\frac{2\chi I_2}{X^{+ 3}}\right)+O(s^3).
\end{align}
%For the last integral one needs
%\beq
%\int_0^L\dd X^+ \chi^n X^{+\,m}=L^{2n+m+1}\frac{\Gamma(n+1)\Gamma(n+m+1)}{\Gamma{(2n+m+2)}}\,.
%\eeq
Finaly, we have
\begin{align}\label{eq:X+integral}
V^{(4)}=&\int_0^L\dd X^+ S_\Omega^{(4)}=\tau^4\pi\left(\frac{1}{24}+\frac12\int_0^L\dd X^+ \frac{(L-X^+)^2}{L^3X^+}\int_0^{X^+} \dd x \int_0^{x} \dd x'\;x'  R\indices{_+_+}(x')+O(L^3)\right)+O(s^3).
\end{align}
where we have used (\ref{propertime}) for $\tau$.\\

In principle, volume for any dimension $n$ can be calculated using our methods demonstrated above.

\end{document}